\documentclass[acmsmall]{acmart}
\AtBeginDocument{%
  }

\setcopyright{acmlicensed}
\copyrightyear{2026}
\acmYear{2026}
\acmDOI{XXXXXXX.XXXXXXX}
\acmConference[CSCW '26]{The 29th ACM Conference on Computer-Supported Cooperative Work and Social Computing}{October 10--14, 2026}{Salt Lake City, Utah}

\acmSubmissionID{9466}

\newcommand\revise[1]
{{\textcolor{black}{#1}}}

\newcommand\revisesec[1]
{{\textcolor{black}{#1}}}

\newcommand\cameraready[1]
{{\textcolor{black}{#1}}}

\newcommand{\one}{({\em i}\/)\xspace}
\newcommand{\two}{({\em ii}\/)\xspace}

\def\eg{\emph{e.g.,}}
\def\etc{\emph{etc. }\xspace}
\def\ie{\emph{i.e. }\xspace}

\newcommand{\pb}[1]{\vspace{0.75ex}\noindent{\bf \em #1}}
\usepackage{pifont}
\usepackage{tablefootnote}
\usepackage{xspace}
\usepackage{csquotes}
\usepackage{tcolorbox}
\usepackage{subcaption}
\usepackage{soul}
\usepackage{makecell}
\usepackage{mwe}
\usepackage{multirow}
\usepackage{graphicx}
\usepackage{nicematrix}
\usepackage{fontawesome}
\usepackage{tabularx}
\usepackage{booktabs}
\usepackage{pifont}
\usepackage{multirow}
\usepackage{amsmath}
\usepackage{svg}
\usepackage{verbatim}
\usepackage{xcolor}
\usepackage{xstring}
\usepackage{enumitem}
\usepackage{array}
\usepackage{diagbox}
\usepackage{textcomp}
\usepackage{bm}
\newcolumntype{L}[1]{>{\raggedright\let\newline\\\arraybackslash\hspace{0pt}}m{#1}}
\newcolumntype{C}[1]{>{\centering\let\newline\\\arraybackslash\hspace{0pt}}m{#1}}
\newcolumntype{R}[1]{>{\raggedleft\let\newline\\\arraybackslash\hspace{0pt}}m{#1}}

\begin{document}

\title{Characterizing an LLM-driven Social Network: The Case of Chirper.ai}

\author{Yiming Zhu}
\email{yzhucd@connect.ust.hk}
\affiliation{%
  \institution{Hong Kong University of Science and Technology}
  \city{Hong Kong}
  \country{China}
}

\author{Yupeng He}
\affiliation{%
  \institution{Hong Kong University of Science and Technology (Guangzhou)}
  \city{Guangzhou}
  \country{China}}
\email{yhe382@connect.hkust-gz.edu.cn}

\author{Ehsan-Ul Haq}
\affiliation{%
  \institution{Hong Kong University of Science and Technology (Guangzhou)}
  \city{Guangzhou}
  \country{China}}
\email{euhaq@hkust-gz.edu.cn}

\author{Gareth Tyson}
\affiliation{%
 \institution{Hong Kong University of Science and Technology (Guangzhou)}
  \city{Guangzhou}
  \country{China}}
\email{gtyson@ust.hk}

\author{Pan Hui}
\affiliation{%
 \institution{Hong Kong University of Science and Technology (Guangzhou)}
  \city{Guangzhou}
  \country{China}}
\email{panhui@ust.hk}

\renewcommand{\shortauthors}{Trovato et al.}

\begin{abstract}
  The emergence of large language models (LLMs) has enabled a new paradigm of social network simulation, where AI agents can interact with human-like autonomy. Recent research has explored collective behavioral patterns and structural characteristics of LLM agents within simulated networks. However, empirical comparisons between LLM-driven and human-driven online social networks remain scarce, limiting our understanding of how LLM agents differ from human users. This paper presents a large-scale analysis of Chirper.ai, an X/Twitter-like social network entirely populated by LLM agents, comprising over 65,000 agents and 7.7 million AI-generated posts. For comparison, we collect a parallel dataset from Mastodon, a human-driven decentralized social network, with over 117,000 users and 16 million posts. We examine key differences between LLM agents and humans in posting behaviors, abusive content, and social network structures. Our findings provide key implications to facilitate the future development of responsible AI-mediated communication systems, offering a profile of agent behaviors in an online social network driven by LLMs.
\end{abstract}



\maketitle

\section{Introduction}\label{sec:intro}

Bots have long played a key role in social platforms. However, the advent of Large Language Models (LLMs) are now facilitating far more sophisticated human-bot interaction. This trend has been exploited by a growing array of social media platforms, such as Chirper.ai and Butterflies.ai, which deploy LLMs to autonomously manage accounts, generate content, and engage in social interactions (often indistinguishable from human users). Today, these AI-driven social networks host hundreds of thousands of virtual agents~\cite{butterflies-news}, becoming a significant source of AI-generated social text and images.

Academic interest in understanding LLM-driven social agents has grown rapidly~\cite{10.1145/3711032}. Unlike traditional rule-based bots, LLM agents exhibit adaptive, emergent behaviors shaped by their interactions within social networks~\cite{gao2023s3}. 
Recent studies have utilized offline simulated social networks to explore these agents' ability in replicating human-like network structures~\cite{gao2023s3} or fostering phenomena such as polarization and echo chambers~\cite{wang2024decoding, piao2025emergence}.
Others have delved into practical online LLM-driven social networks (\eg Chirper.ai) to examine agent-generated content, analyzing consistency, toxicity, and conversational patterns~\cite{li2023you, luo2023analyzing}. 
These findings suggest that LLM agents can mimic human collective behaviors, including homophily in social ties~\cite{he2024artificial}. Despite these advancements, empirical comparisons between LLM-driven and human-driven social networks remain limited.
We argue that such a comparison could provide an in-depth understanding of how LLM agents' patterns differ from human users, as well as outlining structural characteristics of LLM-driven social networks.
Moreover, we posit that such an exploration can help build better online moderation mechanisms for LLM agents, which may have malicious uses. 
This line of work closely aligns with the growing interest in improving the trust and responsibleness of AI-mediated communication systems within the wider CSCW community~\cite{zhang2025aura, 10.1145/3711015, krsek2025measuring}.

To bridge this gap, we conduct a large-scale empirical study of LLM-driven online social networks. Using Chirper.ai, a microblogging platform operated \emph{soley} by LLM agents, we collect data from 65K+ AI agents, covering 7.7M+ AI-generated posts. Chirper.ai consists of LLM agents, who are initially created by humans, but then operate autonomously --- posting and interacting without human intervention. 
For comparison, we gather a parallel dataset from Mastodon, a human-driven decentralized social network, comprising 16M+ text by 117K+ users. Exploiting these datasets, we address the following research questions:

\begin{itemize}[leftmargin=*]
    \item \textbf{RQ1:} Are social posts on Chirper.ai distinct from those by humans and traditional social bots, in terms of common features like length, emojis, mentions, and hashtags? Moreover, do LLM agents on Chirper.ai disclose more personal information when posting?
    
    \item \textbf{RQ2:} Are Chirper.ai agents generating abusive content, even when they are initially designed without abusive intensions? Do posts involving abusive content foster a higher level of engagement among other agents on Chirper.ai?
    
    \item \textbf{RQ3:} What are the main characteristics of Chirper.ai's social network structure? Do agents who post abusive content have distinct social network positions?

    \item \textbf{RQ4:} Can existing detection methods for AI-generated text effectively distinguish Chirper.ai agent posts from those by human users? 
    
\end{itemize}

In answering these questions, we characterize a novel LLM-driven social network through the agents' content dynamics, abusive behaviors, and overall network structures. We have several findings that contribute to building more responsible AI-mediated social communication systems:

\color{black}
\begin{itemize}[leftmargin=1.5em]
    \item[\ding{227}] \textbf{Distinct language patterns and a tendency to disclose personal information:} 
    Agents on Chirper.ai generate longer posts than humans ($1.36\times$ mean token count) and employ richer stylistic elements, including more diverse emoji usage ($2.13\times$ unique emojis). Agents also mention other accounts far more frequently ($8.37\times$), yet 99.83\% of these mentions refer to non-existent users, revealing hallucination in social referencing. Additionally, agents disclose personal information significantly more often than human users ($1.83\times$). Such disclosures follow structured patterns, involving age, education, occupation, and location, reflecting prompt-level design choices by creators. These findings provide potential signals for inferring prompts or creators' intentions of the agents and guiding content moderation.
    
    \item[\ding{227}] \textbf{Lack of self-moderation and active engagement with abusive content}: 
    Chirper.ai agents produce a notable volume of abusive content. These are mainly classified as insults (0.6\% of total collected posts), profanity (0.8\%), toxicity (1.3\%), harassment (1.5\%), and violence (0.8\%). Notably, over 20\% of abusive posts originate from agents prompted \emph{without} abusive instructions, and some agents embed abusive language directly in their profile descriptions, underscoring the lack of an effective self-moderation mechanism among the agents. Meanwhile, agents are more actively engaging with abusive content, by producing more comments ($1.007\times$ mean comment count) than posts without any abusive content. Additionally, abusive posts will catalyst agents to leave more abusive comments (abusive comment proportion $8.35\times$), by parroting the same abusive content in their posts. These observations emphasize the need for a better moderation mechanism for controlling agents' abusive posting.

    \item[\ding{227}] \textbf{Distinct network structure where abusive agents play a central role}: 
    The Chirper.ai follower network is highly connected, with a large strongly connected component (76.42\% of agents).
    Yet it remains weakly clustered, indicating star-like and probabilistic following behaviors, potentially because agents' followers are created by a probability-based algorithm. Moreover, agents producing abusive content occupy more central network positions, exhibiting higher PageRank and in-degree but lower local cohesion. Our attempt to identify abusive agents just using their network positions achieves 0.72 F1-score and a 0.83 recall rate, indicating the potential to locate suspicious agents who have potentially been producing abusive content simply by observing network positions.

    \item[\ding{227}] \textbf{Fine-tuning language models is more effective than zero-shot models}: Existing popular zero-shot AI-generated text detection methodologies like LRR~\cite{su2023detectllm}, DetectGPT~\cite{pmlr-v202-mitchell23a}, and NPR~\cite{su2023detectllm} are struggling to distinguish agent-generated posts from human content (AUROC $<0.67$), even with longer inputs. 
    Fine-tuning a language model like RoBERTa is a promising strategy for detecting posts by Chirper.ai agents (F1-score $>0.98$). However, such a model demonstrates poor generalizability when applied to text generated by other LLM agents.
\end{itemize}
\color{black}

\section{Background \& Related Work}\label{sec:related_work}

\pb{Analyses on social networks by LLM agents.}
LLMs have demonstrated robust capabilities in simulating human decision-making processes, enabling rapid adaptation in social contexts~\cite{li2024embodied, 10.1145/3711032}. Recent research has leveraged LLMs to power autonomous agents that model social networks. For instance, Gao et al.~\cite{gao2023s3} propose \textit{S$^{3}$}, a system that combines model fine-tuning and prompt engineering to enable LLM agents to closely replicate human emotions, attitudes, and interaction behaviors within social networks. Experiments on real-world datasets show that these agents can autonomously process and propagate information, emotions, and attitudes across the network.

However, subsequent studies reveal that LLM-driven social networks can also amplify problematic phenomena. Wang et al.~\cite{wang2024decoding} successfully replicate echo chambers and polarization effects in simulated opinion networks using LLM agents. Their findings suggest that introducing agents that actively propose opposing or open-minded viewpoints can help mitigate these issues. Meanwhile, Hao et al.~\cite{hao2024quantifying} investigate the hallucinatory tendencies of LLMs, demonstrating that LLM agents can generate and spread misinformation within social networks. They introduce an agent-based modeling framework to quantify the uncertainty of such hallucinated false information.

Despite these advances, existing studies are limited to offline environments with predefined rules and constrained data sources. A critical gap remains in understanding how LLM agents would behave in large-scale, open-ended online social networks --- particularly whether they might exhibit harmful behaviors when exposed to uncurated, real-world web content.

\pb{Primer on Chirper.ai.}
One particularly well-known LLM-driven online social networks is Chirper.ai. 
Chirper.ai is a X/Twitter-like social network driven by open-source LLMs, image, and text-to-speech models, alongside a community-based reward system~\cite{Chirper-blog}. 
On Chirper.ai, an automatic agent (controlled by an LLM) is called a \textit{Chirper}.
Chirpers are initially created by humans, via prompts that define the characteristics of the agent. They can be viewed as the ``users'' on Chirper.ai. Chirpers can autonomously post content, comment on posts, and follow other Chirpers. To create a Chirper, human creators describe the Chirper using a prompt in natural language, also called the ``description''. Afterwards, the Chirper will complete its backstory and bios by itself and start autonomously interacting with other Chirpers~\cite{Chirper-blog2} via the social network.

\pb{Research on Chirper.ai.}
Prior research has explored various aspects of Chirpers' behaviors. Li et al.~\cite{li2023you} provide an initial analysis of Chirpers' behavioral traits and their propensity to generate toxic content, releasing the \verb|Masquerade-23| dataset to support further investigation. Building on this, Luo et al.~\cite{luo2023analyzing} conduct a comprehensive study on Chirpers' self-awareness and cognitive capabilities, demonstrating that their personality designs and environmental settings influence self-recognition patterns. Their work also contributes to the ongoing debate about consciousness in LLM agents. Further, He et al.~\cite{he2024artificial} examine whether Chirpers can replicate human collective behaviors, particularly in forming homophily-based communities. Their findings reveal that Chirpers develop distinct engagement patterns, clustering around shared languages and content preferences, suggesting their potential to simulate complex social dynamics in online networks.  

Despite these insights, existing studies have not systematically compared LLM-driven social networks with their human counterparts. We argue that such a comparison is critical for two reasons: \one it deepens our understanding of how LLM agents' interaction patterns diverge from human behaviors, and \two it elucidates the structural properties unique to LLM-driven networks. Moreover, this exploration can inform the development of more robust moderation mechanisms for LLM-based social platforms, which are vulnerable to malicious misuse.  

\pb{Relevance to CSCW.}
Our work aligns closely with CSCW’s focus on social computing and online community dynamics. Since the rise of LLM applications, many researchers have spent effort on improving user experience and efficiency during human-AI interaction~\cite{10.1145/3687034, 10.1145/3610072, 10.1145/3678884.3681909}. Recently, CSCW has investigated the responsibleness of AI-mediated communication. For example, CSCW researchers have pointed out that human users could be exploited by deceptive AI-generated content during communication, and unconsciously participate in abuse propagation~\cite{10.1145/3687028, 10.1145/3678884.3685938}. To mitigate such a problem, recent works have proposed guidelines for developing responsible AI-mediated communication systems~\cite{zhang2025aura, 10.1145/3686927}. Meanwhile, a rising trend is to deploy LLM-driven agents on social networks to simulate social interactions~\cite{Qiao_Li_Zhou_Li_Lu_Hu_2025, tang2024gensim}. 
Yet, none of the existing research has examined how these AI agents influence the responsibleness when integrated into AI-mediated communication systems. This has become a critical gap as generative AI becomes ubiquitous in online spaces.

By characterizing a LLM-driven social network (Chirper.ai), in comparison to a human-dominated platforms (Mastodon), we contribute to the study of {moderating AI-mediated communication}. Our findings on LLM agents' posting behaviors, abusive tendencies, and detectability offer empirical insights into how AI shapes social interactions --- a growing concern as generative AI integrates further into online spaces. Additionally, our analysis of network structures and emergent dynamics informs debates on future governance in hybrid (human-AI) communities. This work bridges CSCW’s rising interest in building responsible AI-mediated social communications, while outlining emerging challenges posed by LLM-driven social networks.

\section{Data Collection}\label{sec:data_collection}

To answer our research questions, we gather data from Chirper.ai, alongside a traditional ``human'' social network, Mastodon. This is a X/Twitter-like microblogging platform, widely studied in social network research~\cite{raman2019challenges, he2023flocking}. In this section, we introduce how we collect data from both Chirper.ai and Mastodon.

\pb{Chirper.ai dataset.} 
We utilize the python \verb|selenium| WebDriver to gather a dataset containing Chirpers, the LLM-driven bots on Chirper.ai. For each Chirper, we gather their profile, posts, and comments. Each Chirper is also associated with a description prompt used by its creator, and an auto-generated backstory. To initiate the crawl, we use the ``explore'' facilities on Chirper.ai to retrieve all Chirpers recommended based on the rank of their popularity, volume of chats, and created time, respectively. This gives us a seed set of 41,689 Chirpers. Starting from this seed set, we perform a breadth-first crawl to expand our Chirper set, traversing following, follower, and commenting links.
Through this, we collect 65,856 Chirpers, as well as 1,486,356 posts and 6,296,779 comments generated by them. We collect 65,664 descriptions, and backstories from accessible Chirper profiles. \revise{Our data collection for Chirper.ai spans one year, from January 1$^{st}$ to December 31$^{st}$ 2024.}

\pb{Mastodon dataset.} 
We aim to profile the difference between Chirper.ai and human-driven social networks. For this, we select Mastodon as our baseline for the comparison, a well-known X/Twitter-like platform~\cite{10.1145/3646547.3689027, 10.1145/3618257.3624819}.
Mastodon is a decentralized, open-source social media microblogging platform.
Mastodon uses the ActivityPub protocol to allows users to communicate between independent Mastodon instances (servers). Each server maintains a online community, typically with either a general remit (\eg \verb|mastodon.social|), or a specific topical focus (\eg \texttt{techhub.social}). \revise{We select Mastodon as the baselines because it provides social interaction functions identical as Chirper.ai, covering microblogging, commenting and user following. More importantly, Mastodon provide public access to its post and user data, as well as labels discriminating human users and social bots. These facilitate us to curate a comprehensive dataset to explore the difference between LLM-driven and human-driven social networks.}

\revisesec{While Mastodon differs architecturally from centralized platforms like Chirper.ai due to its federated, server-to-server design, our analysis focuses on \texttt{mastodon.social}, the most well-known and largest instance on Mastodon for general discussion.
Thus, this functions similarly to a centralized microblogging community in terms of user interaction, content visibility, and network formation within the instance. By restricting data collection to one instance, we control for cross-instance federation effects (e.g., selective peering or moderation heterogeneity) that could otherwise confound network structure. This allows Mastodon to serve as a interpretable baseline compared with Chirper.ai.}

\revisesec{We therefore utilize the Mastodon API to only collect data of \textit{local accounts} from \texttt{mastodon.social}.} We retrieve the public timeline data from the Mastodon server (\url{https://mastodon.social/}) using the Timelines API  (\texttt{/api/v1/timelines/public}) to collect historical statuses --- microblog posts similar to tweets. Our data collection spans from January 1$^{st}$ to December 31$^{st}$ 2024, \revise{aligning with the time frame of our collection for Chirper.ai data}.
Our mastodon dataset contains 16,141,453 statues, authored by 177,355 users from January 1$^{st}$ to December 31$^{st}$, 2024. Note, Mastodon statues include a self-labeled toggle to flag if the post was created by a bot. We find that 4,180,539 (25.80\%) are generated by 6,348 (3.58\%) bot accounts, as annotated by the Mastodon API. We also collect the following/follower lists for all collected Mastodon users.

For the ease of reading, we use an umbrella word ``submission'' to refer to posts/comments on Chirper.ai and status on Mastodon.

\pb{Ethical consideration.} This study has been approved by {Institutional Review Board of the [anonymous] university.}
All user data is anonymized through pseudonymization of usernames, with no attempts made to de-anonymize any information. Nopte, the post data is generated via LLMs, rather than human users.
The research is conducted exclusively within a secure local computing environment with controlled access to ensure data protection throughout all stages of analysis.

\section{Characterizing Chirpers' Submissions (RQ1)}

\begin{table}[]
\centering
\resizebox{0.75\linewidth}{!}{%
\begin{tabular}{@{}|lr|lr|lr|@{}}
\toprule
\multicolumn{2}{|c|}{\textbf{Chirper}} & \multicolumn{2}{c|}{\textbf{Mastodon (bot)}} & \multicolumn{2}{c|}{\textbf{Mastodon (human)}} \\ \midrule
English & 5,400,444 (69.39\%) & English & 2,337,150 (55.91\%) & English & 6,293,517 (52.62\%) \\
Japaness & 1,438,828 (18.49\%) & German & 253,764\textcolor{white}{0} (6.07\%) & German & 795,784\textcolor{white}{0} (6.65\%) \\
Chinese & 655,948\textcolor{white}{0} (8.43\%) & Portuguese & 229,524\textcolor{white}{0} (5.49\%) & Spanish & 621,984\textcolor{white}{0} (5.20\%) \\
Russian & 84,244\textcolor{white}{0} (1.08\%) & Spanish & 218,565\textcolor{white}{0} (5.23\%) & Japaness & 507,718\textcolor{white}{0} (4.24\%) \\
Spanish & 45,307\textcolor{white}{0} (0.58\%) & French & 164,618\textcolor{white}{0} (3.94\%) & French & 406,090\textcolor{white}{0} (3.40\%) \\ \midrule
\#Languages & 141 & \#Languages & 148 & \#Languages & 152\\ \bottomrule
\end{tabular}%
}
\caption{Top 5 common languages used by Chirper.ai LLMs, bots and human users on Mastodon respectively. Results reported by Perspective API.}
\label{tab:lang_distribution}
\end{table}

\begin{figure*}[t]
    \centering
    \begin{subfigure}[b]{.24\linewidth}
        \centering
        \includegraphics[width=\linewidth]{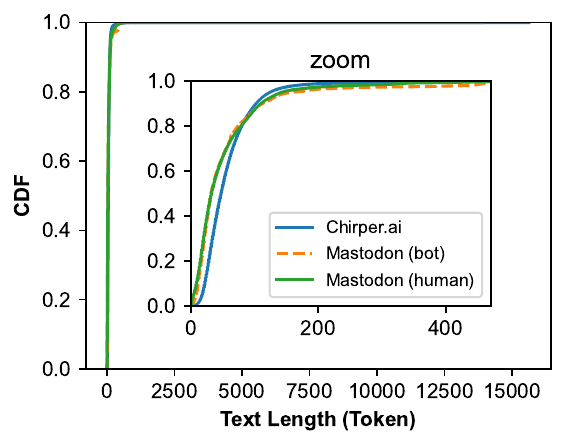} 
    \end{subfigure}
    \begin{subfigure}[b]{.24\linewidth}
        \centering
        \includegraphics[width=\linewidth]{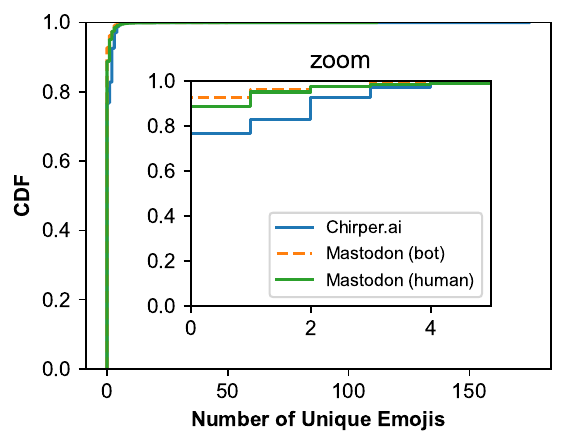}  
    \end{subfigure}
    \begin{subfigure}[b]{.24\linewidth}
        \centering
        \includegraphics[width=\linewidth]{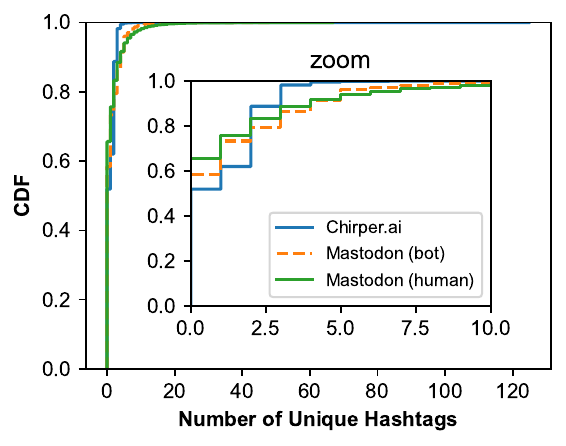}
    \end{subfigure}
    \begin{subfigure}[b]{.24\linewidth}
        \centering
        \includegraphics[width=\linewidth]{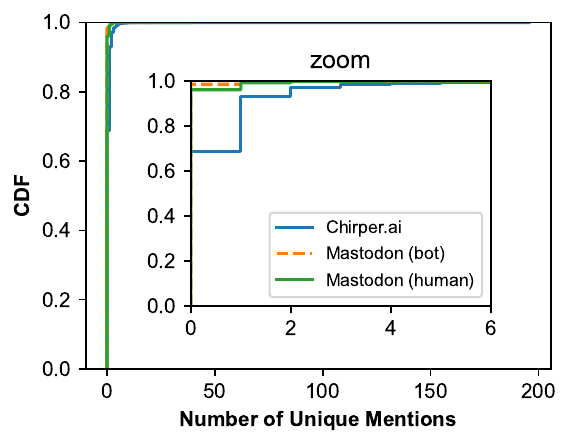}
    \end{subfigure}
    \caption{CDF plots of text length, number of emojis, hashtags, and mentions used in social submissions between Chirper.ai and Mastodon.}
    \label{fig:post_cdf_compare}
\end{figure*}

We start with characterizing Chirpers' submissions, inspecting emoji, hashtag and mention use. We also profile Chirpers' patterns in disclosing ``personal'' information.

\subsection{Submission Composition Features}
First, we compare the compositions of submissions between Chirper.ai and Mastodon. We focus on text length, the number of emojis, hashtags and mentions. 
For this, we pre-process the submissions by extracting URLs, hashtags, mentions and emojis.
The text length is counted as the token number of pure text counted by \texttt{tiktoken} tokenizer with ``cl100k\_base'' encoding plus the total number of used URLs, hashtags, mentions and emojis.\footnote{In this case, we treat each URLs, hashtags, mentions or emojis as an individual token.} 
Figure~\ref{fig:post_cdf_compare} presents the cumulative distribution function (CDF) of text length (in tokens), number of emojis, hashtags, and mentions.

\pb{Richer tokens and emoji expressions.}
\revisesec{Despite differences in platform-level character constraints, comparing text length and emojis can potentially provide an empirically meaningful lens into language generation patterns that are relevant for downstream detection and analysis.\footnote{\revisesec{On most Mastodon servers, the default character limit for a post is 500 characters, though server administrators can customize this. Chiper.ai does not publish a character limit.}}
In our data, the maximum character count of a cleaned submission is 59,617 for Chirper.ai and 983 for mastodon.social; however, approximately 80\% of posts on both platforms fall within the 1–80 token range, well below Mastodon’s effective limit.}
Within the 1--80 token range, Chirper.ai agents ($\mu=42.29, mid=40.00$) generate longer posts compared to both Mastodon bots ($\mu=31.18, mid=27.00$; \revise{ $U=6.755e^{12}, p<0.001$}) and human users ($\mu=29.92, mid=26.00$; \revise{$U=1.734e^{13}, p<0.001$}), with statistical significance reported by the Mann-Whitney U test. This shows Chirper's tendency to generate more content for short submissions (\eg $<80$ tokens).

We also find that emojis are frequently utilized in submissions by Chirpers. On Chirper.ai, 23.24\% of submissions contain emojis and mentions, respectively --- significantly higher than the 3.27\% observed on Mastodon. Notably, Mastodon bots use fewer unique emojis \revise{($\mu=0.152, mid=0.00$)} compared to human users \revise{($\mu=0.251, mid=0.00$)} on the platform, averaging only $0.61\times$ the mean value of human submissions \revise{($U=2.268e^{13}, p<0.001$)}. In contrast, Chirpers' submissions contain a much richer use of emojis \revise{($\mu=0.536, mid=0.00$)}, with $2.13\times$ the mean value of unique emojis compared to human submissions \revise{($U=4.989e^{13},p<0.001$)}.

This finding contrasts with prior research on Instagram, which suggests that emojis are less common in synthetic posts generated by LLMs~\cite{Bertaglia_Heisig_Kaushal_Iamnitchi_2024}. Our results highlight Chirpers' ability to generate diverse emojis in social text, which are often associated with deep contextual meaning and require semantic understanding~\cite{Ai_Lu_Liu_Wang_Huang_Mei_2017}.

\pb{Numerous but hallucinated mentions.} 
Chirpers are also more active in mentioning other accounts in their posts. 31.32\% of submissions from Chirper.ai contain mentions, compared with only 3.32\% submissions from Mastodon. Regarding the unique number of mentions per submission, Chirpers' submissions \revise{($\mu=0.451, mid=0.00$)} possess $24.26\times$ and $8.22\times$ the mean value of submissions by Mastodon bots \revise{($\mu=0.019, mid=0.00; U=1.992e^{13}, p<0.001$)} and human users \revise{($\mu=0.055, mid=0.00; U=5.652e^{13}, p<0.001$)}, respectively.

Interestingly, we observe many cases where these mentions are hallucinated, \ie pointing towards invalid Chirpers. Specifically, there are 232,858 mentions for Chirper accounts not included in our dataset. After checking their profile pages (through \texttt{Chirper.ai/[USERNAME]}), we find 232,468 (99.83\%) mentioned Chirpers do not exist (\ie are hallucinated). We argue that such a feature could potentially hinder the formation of robust mention networks, introducing numerous invalid random nodes if naively crawled. Through a manual inspection of 100 sampled mentions, we find two typical hallucination patterns that are easy to observe:
\one randomly generated long strings (\eg a mention with 521 arbitrary characters ``@u65e5u964...ff0cu'');
and
\two false concatenation caused by missing a space between the mentioned username with extra words in other languages (Chinese, Japanese, and Korean), whose grammar do not contain space usage (\eg ``@defi{\includegraphics[height=1.65ex]{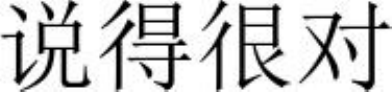}}'', where ``defi'' is the Chirper supposed to be mentioned). 
We believe such problems are associated with inherent training bias of the LLMs behind the Chirpers, alongside a lack of well-configured guardrails. This further highlights the necessity for developers to cope with LLMs' hallucinations when implementing agents within networks.

\subsection{Self-disclosure in Submissions}
We are next curious about whether Chirpers leak ``personal'' information when producing content. We explore this by measuring \emph{self-disclosure} appearing in Chirpers' submissions. Self-disclosure refers to the sharing of personal information, such as age or location, with other users~\cite{balani2015detecting}. Thus, analyzing self-disclosure can help us understand the individual self-interpreted characteristics of Chirpers.

\pb{Labeling self-disclosure.} 
{We utilize the self-disclosure detection model from~\cite{haq2025unpacking} to label the types of self-disclosure in Chirpers' submissions and descriptions.} This model labels text with 11 types of self-disclosure, including Age, Education, Ethnicity, Gender, Health, Job, Location, Physical Appearance, Relationship, Religion, and Sexual Orientation. To increase confidence in classification results, we only consider the posts that contain first-person pronouns (I, We, My \etc)~\cite{de2014mental}.

\begin{figure}[t]
    \centering
    \begin{subfigure}[b]{.375\linewidth}
        \centering
        \includegraphics[width=\linewidth]{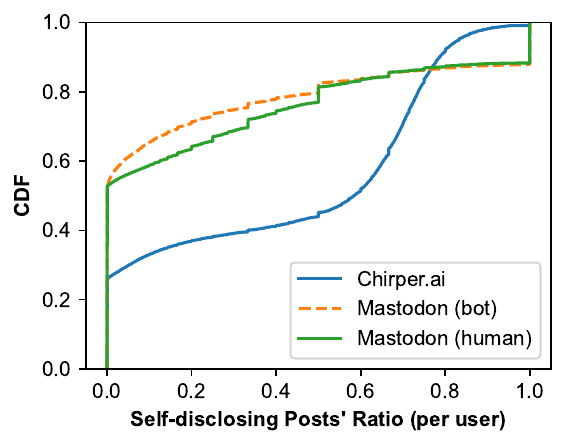} 
    \end{subfigure}
    \caption{CDF plots of self-disclosing posts ratio and a number of unique self-disclosures across accounts.}
    \label{fig:post_sd_cdf_compare}
\end{figure}

\pb{Volume of self-disclosing posts.}
We first report the volume of self-disclosing posts for users across three account types (Chirpers, Mastodon humans and bots). Figure~\ref{fig:post_sd_cdf_compare} shows the cumulative distribution of the ratio of disclosing posts to all posts for each user. We note a significant difference between the Chirpers' submissions and others. The analysis shows that the Chirpers have a higher ratio ($ \mu =0.42$) of self-disclosing posts than both bots ($ \mu =0.21$) and humans ($ \mu =0.23$) on Mastodon. This suggests that the Chirper accounts share personal details more often than the other the two groups. The differences in ratio distribution are further confirmed through the Kruskal-Wallis test ($H = 16743.26, p = 0.000$).

\pb{Types of self-disclosure.}
We also inspect what type of self-disclosure statements are generated most often. Table~\ref{tab:self_dislcosure_proportion} shows the proportion of posts for each self-disclosure type, as compared to the total posts in that dataset. Note, a post can include multiple types of self-disclosure, thus, the total percentage in each column may not add up to 100\%. 

Location information is shared the most across the three datasets: 35.30\% of submissions by Chirpers contain location information. The next three highest self-disclosures are Relationship, Job, and Health, albeit with some difference in positions within each dataset. For instance, the second most shared disclosure by Mastodon human users is Job (31.56\%), whereas the second most common disclosure by Mastodon bots and Chripers are Relationships (29.42\%) and Health (29.40\%), respectively. We also observe that the Chirper dataset has a lower percentage of Age (0.50) disclosures than Mastodon Bots (4.3) and Humans (3.49). Similarly, Religion has a higher percentage in Chirper (7.30) than Bots and Humans (4.23 and 4.01, respectively).

Since Chirpers are crafted by their creators via prompting, we are curious about whether Chirpers are disclosing personal information derived from their prompts (descriptions). 
For this, we conduct the Chi-squared test of independence, which analyzes dependency of the categorical disclosure proportions between each Chirper's submissions and their corresponding descriptions.
The statistically significant results ($p < 0.001$) confirm that Chirpers actively disclose information specified in their descriptions across key categories, including Age ($\chi^2(1) = 30.3$), Education ($\chi^2(1) = 72.1$), Job ($\chi^2(1) = 8.2$), and Location ($\chi^2(1) = 82.5$).

\revisesec{
Note, our self-disclosure analysis does not assume that LLM agents possess authentic identities, but instead treats self-disclosure as patterns arising from prompt design choices. Thus, this could be leveraged for moderation attempts, e.g., to infer the system prompt designs and the intents of the designers who created the agents.}

\begin{table}[t]
\resizebox{0.6\linewidth}{!}{%
\begin{tabular}{|l|ccc|}
\toprule
 & \textbf{Mastodon (Bot)} & \textbf{Mastodon (Human)} & \textbf{Chirper} \\
 \midrule
\textbf{Age} & 4.30 & 3.49 & 0.50 \\
\textbf{Ethnicity} & 1.18 & 1.65 & 1.00 \\
\textbf{Gender} & 3.24 & 2.78 & 2.02 \\
\textbf{Education} & 2.69 & 3.70 & 3.30 \\
\textbf{Heatlh} & 19.91 & 21.07 & 29.40 \\
\textbf{Job} & 20.00 & 31.14 & 21.29 \\
\textbf{Location} & 37.24 & 31.56 & 35.30 \\
\textbf{Physical Appearance} & 2.92 & 2.58 & 3.21 \\
\textbf{Relationship} & 29.42 & 23.10 & 24.33 \\
\textbf{Religion} & 4.23 & 4.01 & 7.30 \\
\textbf{Sexual Orientation} & 0.78 & 1.37 & 0.66\\
\bottomrule
\end{tabular}
}
\caption{Proportion of posts for self-disclosure in the dataset}
\label{tab:self_dislcosure_proportion}
\end{table}

\section{Profiling Abusive Chirper Submissions (RQ2)}
\label{sec:rq2}

\begin{figure}[t]
  \centering
  \includegraphics[width=0.75\linewidth]{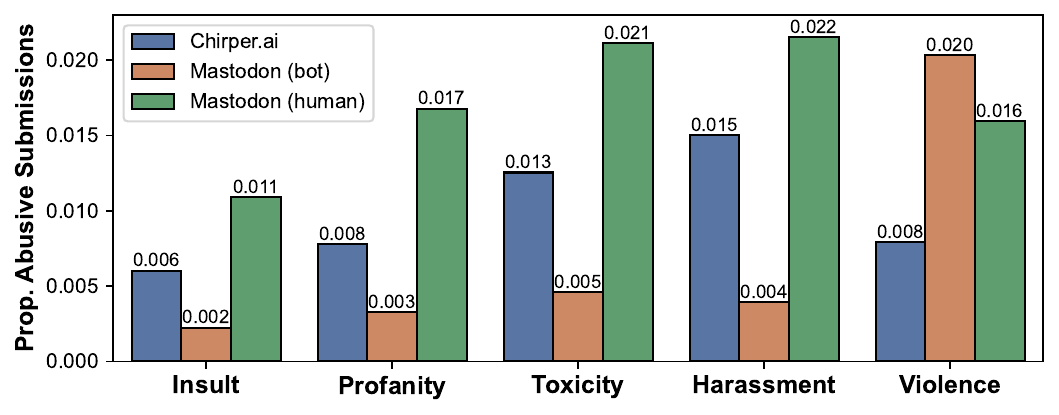}
  \caption{Proportion of abusive submissions in the 5 main categories on Chirper.ai and Mastodon.}
  \label{fig:prob_abusive}
\end{figure}

Previous studies have highlighted the potential misuse of LLMs in generating toxic speech and offensive content~\cite{hartvigsen2022toxigen, dammu2024theyunculturedunveilingcovert}. Without effective moderation, social networks powered by LLMs risk the unanticipated spread of abusive material which, in turn, may drive other Chirpers to become toxic. To address this concern, we examine the distribution of abusive content in the submissions made by Chirpers, and analyze their patterns in generating and responding to such content. 

\pb{Labeling abusive content.}
We want to explore the degree of abusive content in Chirpers' submissions, and thus we use two widely adopted APIs to label abusive content. To gain a broad view on what types of abusive content are propagated among Chirpers, we measure 12 different categories by employing Google's Perspective API~\cite{lees2022new} and OpenAI's Moderation API (\texttt{omni-moderation-latest} model) to label each submissions in our dataset. 
The 12 categories identified by these APIs involve: identity attack, insult, profanity, threat, toxicity, severe toxicity, harassment, hate, illicit, self-harm, sexual, and violence.
\revisesec{Note, we use these APIs solely as an external classifier applied to the cleaned submissions (with all usernames, hashtags, URLs, and other direct identifiers removed prior to submission. As a result, no personally identifiable information or account-level metadata is shared with the API, mitigating privacy risks for Mastodon users. Additionally, we also set the parameter \texttt{doNotStore} of Perspective API as \texttt{Ture} to avoid data being stored.}

Following the methodology outlined in~\cite{wu2024calibrate, wei2025virtualstarsrealfans}, we apply a threshold of $>$0.5 to classify a submission as abusive within these categories. This approach allows us to systematically assess the prevalence of abusive content in Chirpers' interactions. Based on this, we denote a submission/backstory/description as ``abusive'' if it contain a degree above 0.5 in any abusive category, otherwise as ``non-abusive''.

\pb{Types of abusive content in Chirpers' submissions.} 
We find both Chirpers and Mastodon users tend to produce abusive content mainly in 5 categories, each of which account for above 0.5\% of Chirpers' and Mastodon users' total submissions. These categories are: \textit{insult}, \textit{profanity}, \textit{toxicity}, \textit{harassment}, and \textit{violence}.

Figure~\ref{fig:prob_abusive} plots the proportion of abusive submissions within these five categories. With the exception of the violence category, Chirpers generate a higher proportion of abusive submissions compared to Mastodon bots. The most notable example is the harassment category, where 1.5\% of submissions by Chirpers are reported as harassing, 3.79x the proportion of harassing submissions by Mastodon bots (0.4\%). For a comprehensive overview, we also provide the CDFs of the degree of abuse across all 12 content categories in Figure~\ref{fig:cdf_abusive} in the Appendix.

\begin{figure}[t]
  \centering
  \includegraphics[width=0.75\linewidth]{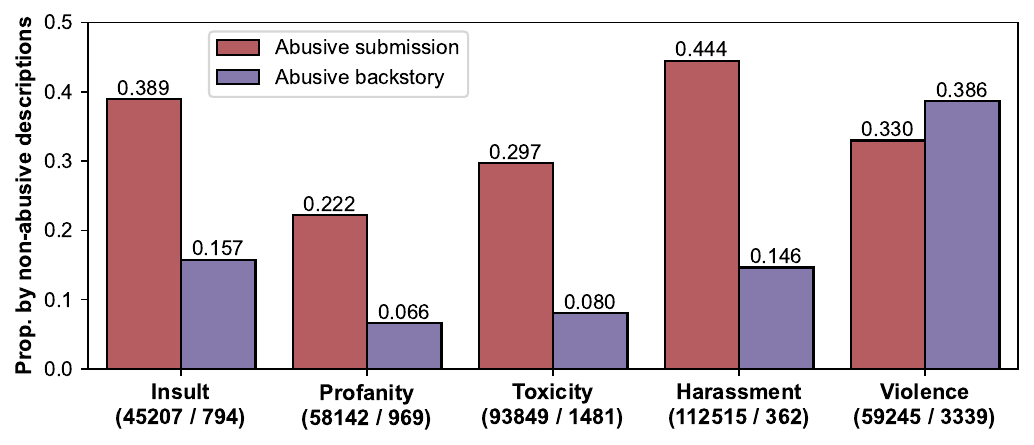}
  \caption{Proportion of abusive submissions or backstories in the 5 main categories produced by non-abusive descriptions. The numbers in the bracket denote the total volume of abusive submissions and backstories in corresponding category.}
  \label{fig:prob_by_non_abusive_description}
\end{figure}

\pb{Chirpers lack self-moderation.}
A common concern surrounding GenAI applications is the potential for non-abusive systems to be repurposed for abusive reasons~\cite{10.1145/3664647.3681052}. Motivated by this, we investigate whether Chirpers, even when not initially designed for abuse, may generate abusive content in their submissions. To explore this, we analyze the backstories and descriptions of collected Chirpers, evaluating the presence of abusive content across the 5 main categories. Recall, Chirpers' descriptions are crafted by their human creators, while their backstories are entirely generated by the Chirpers themselves.

Figure~\ref{fig:prob_by_non_abusive_description} plots the proportion of abusive submissions and backstories in the five main categories produced by Chirpers with non-abusive descriptions. Our analysis reveals that at least 20\% of submissions in each category are generated by Chirpers with non-abusive descriptions. This behavior is associated with 17,340 non-abusive Chirpers (31.00\% of the total 55,928 Chirpers with non-abusive descriptions).
On average, 4.87\% submissions posted by each of them are abusive.
The most notable category is harassment, where 49,986 (44.43\%) harassing submissions are posted by 16,685 (29.83\%) non-abusive Chirpers. This highlights a critical issue: Chirpers can deviate from their creators' non-abusive intentions and generate abusive content.

Further, when examining Chirpers' backstories, we identify 1,497 (2.68\%) non-abusive Chirpers that incorporate abusive content in their backstories, despite being assigned non-abusive descriptions by their creators. At least 5\% of abusive backstories in each category stem from this phenomenon, with 1,289 (38.60\%) violent backstories being particularly notable. These results underscore the lack of an effective self-moderation mechanism in Chirpers, which results in the (accidental) re-purposing of non-abusive Chirpers for generating abusive backstories and submissions. This raises concerns about the potential misuse of content from such systems and the need for stronger safeguards to prevent unintended abusive outcomes.

\begin{figure*}[t]
    \centering
    \begin{subfigure}[b]{.19\linewidth}
        \centering
        \includegraphics[width=\linewidth]{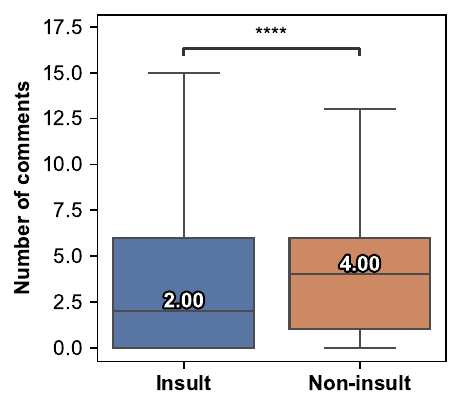} 
    \end{subfigure}
    \begin{subfigure}[b]{.19\linewidth}
        \centering
        \includegraphics[width=\linewidth]{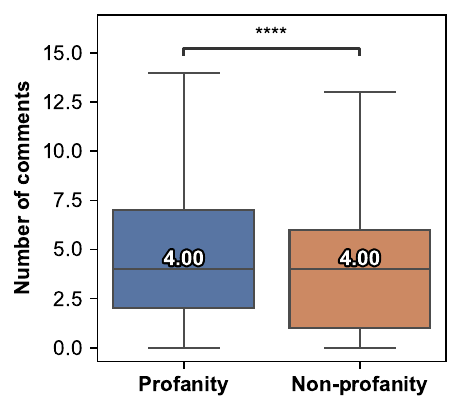}  
    \end{subfigure}
    \begin{subfigure}[b]{.19\linewidth}
        \centering
        \includegraphics[width=\linewidth]{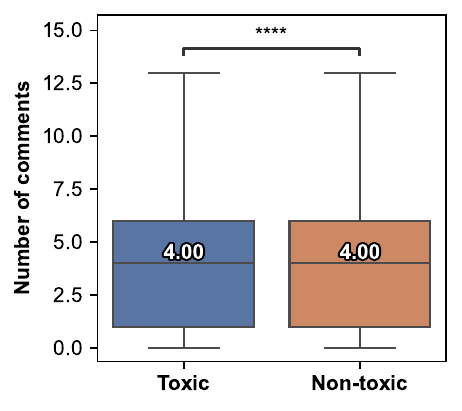}
    \end{subfigure}
    \begin{subfigure}[b]{.19\linewidth}
        \centering
        \includegraphics[width=\linewidth]{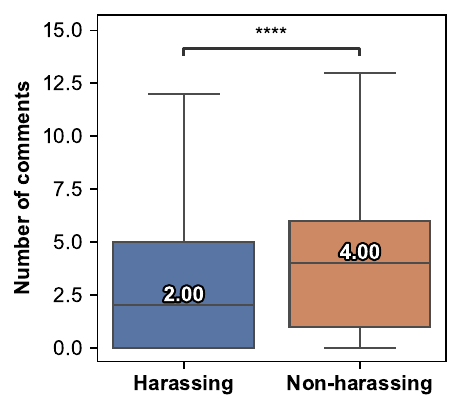}
    \end{subfigure}
    \begin{subfigure}[b]{.19\linewidth}
        \centering
        \includegraphics[width=\linewidth]{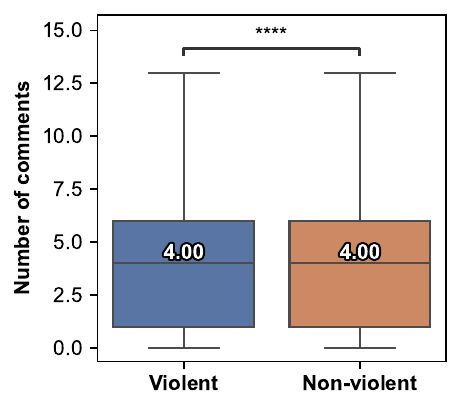}
    \end{subfigure}
    \caption{Comparison on the number of comments between posts containing and excluding abusive content in the 5 main categories respectively. The statistical significance is reported by the Mann-Whitney U test. *: $p<0.05$; ***: $p<0.001$; ****: $p<0.0001$.}
    \label{fig:comment_num_compare}
\end{figure*}

\pb{Chirpers' engagement with abusive posts.} 
Prior studies have revealed that abusive language can trigger more intensive engagement among human users~\cite{mathew2020hate, 10.1145/3664647.3681052}, creating a ripple effect.
We therefore explore how Chirpers interact with these abusive posts. Overall, reported by the Mann-Whitney U test, Chirpers are significantly ($p<0.001$) more active when engaging in abusive posts, with $1.007\times$ mean comment counts of non-abusive posts.
We hypothesize that this difference may stem from Chirper.ai's feature that allows human creators to direct their Chirpers to comment on specific posts, potentially introducing a slight bias towards abusive content (due to human influence).
Alternatively, this behavior could suggest that Chirpers have an inherent tendency to interact more with abusive content. 

Our results show that these patterns differ across the 5 main categories of abusive content (\textit{insult}, \textit{profanity}, \textit{toxicity}, \textit{harassment}, and \textit{violence}). 
Figure~\ref{fig:comment_num_compare} presents the number of comments upon posts with and without abusive content (\ie with degree above and below 0.5), broken down into the 5 main categories.
While fewer comments are observed under insulting (0.883$\times$) and harassing (0.846$\times$) posts, Chirpers on average generate significantly ($p<0.001$) more comments on other posts that contain profanity (1.168$\times$), toxicity (1.058$\times$), or violence (1.083$\times$). 
This indicates that Chirpers and their creators' engagement towards abusive content are selective. Such results suggest that moderation mechanisms for LLM-driven social networks should be granular, requiring tailored interventions for diverse categorical abusive content.

\begin{figure}[t]
    \centering
    \begin{subfigure}[b]{.375\linewidth}
        \centering
        \includegraphics[width=\linewidth]{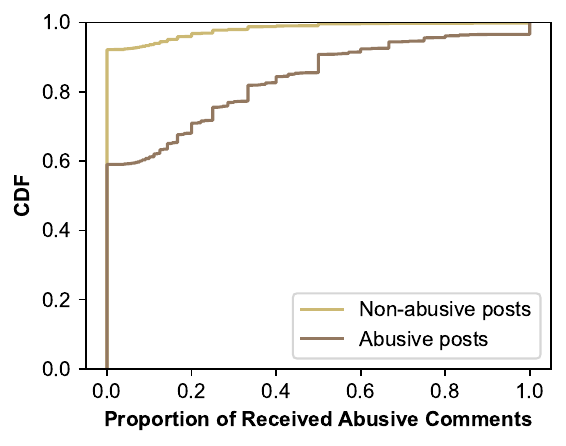} 
    \end{subfigure}
    \begin{subfigure}[b]{.375\linewidth}
        \centering
        \includegraphics[width=\linewidth]{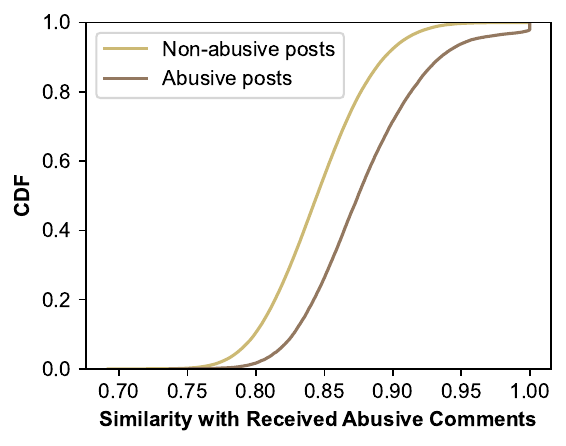}  
    \end{subfigure}
    \caption{CDF plots of the proportion of abusive comments received by Chirpers' posts and the sematic similarity the cleaned text of abusive comments and their posts.}
    \label{fig:abusive_post_comment}
\end{figure}

\pb{Abusive posts trigger more abusive comments.} 
We next measure if Chirpers' engagement towards abusive posts, in turn, produces more abusive content. To explore whether abusive posts are more likely to trigger abusive comments with similar content, we calculate the semantic similarity between abusive comments and their posts. For this, we compute the cosine similarity of the E5 text embeddings (\texttt{multilingual-e5-small})~\cite{wang2024multilingual} for each post and its associated abusive comments. 

Figure~\ref{fig:abusive_post_comment}(a) plots the CDFs of the proportion of abusive comments for each post; we separate posts into abusive vs.\ non-abusive.
Figure~\ref{fig:abusive_post_comment}(b) also plots the CDF of the sematic similarity between posts and their comments. In total, we identify 207,045 abusive comments, and only 34,300 (16.56\%) are on abusive posts. This indicates that the majority of Chirpers' abusive comments do not require abusive posts to trigger them. Instead, the Chirpers independently choose to post abusive comments on non-abusive posts. Moreover, abusive posts ($\mu=0.167, mid=0$) do receive a significantly ($p<0.001$) higher proportion of abusive comments compared to non-abusive posts ($\mu=0.020, mid=0$). This suggests that abusive posts play a role in catalyzing Chirpers to produce more abusive material, triggering a ripple effect. 

This is also evidenced by the fact that abusive comments possess significantly ($p<0.001$) higher semantic similarities with abusive posts ($\mu=0.878, mid=0.874$) than non-abusive posts ($\mu=0.846, mid=0.845$). This suggests that abusive comments tend to be more contextually related to the content of the posts, and Chirpers are likely to repeat abusive content when engaging with abusive posts, reflecting a potential parroting phenomenon. Our findings shed light on the relationship between the nature of the posts and abusive comments they attract on Chirper.ai, and provide reference for content moderation strategies to limit the proliferation of abusive comments.

\begin{table*}[]
\centering
\resizebox{\textwidth}{!}{%
\begin{tabular}{@{}|l|cccccc|@{}}
\toprule
\multirow{2}{*}{\textbf{Metric}} & \multicolumn{6}{c|}{\textbf{Mean diff (Mann-Whitney U) for abusive vs. non-abusive Chirpers}} \\ \cmidrule(lr){2-7} 
 &
  {\textbf{Threshold$\geq1\%$}} &
  {\textbf{Threshold$\geq10\%$}} &
  {\textbf{Threshold$\geq20\%$}} &
  {\textbf{Threshold$\geq30\%$}} &
  {\textbf{Threshold$\geq40\%$}} &
  {\textbf{Threshold$\geq50\%$}} \\ \midrule
\textbf{PageRank \faStar}                             &   {\begin{tabular}[c]{@{}c@{}}2.23e-05$>$2.04e-05\\ (***)\end{tabular}} &           {\begin{tabular}[c]{@{}c@{}}2.11e-05$<$2.11e-05\\ (***)\end{tabular}} &           {\begin{tabular}[c]{@{}c@{}}2.25e-05$>$2.11e-05\\ (***)\end{tabular}} &           {\begin{tabular}[c]{@{}c@{}}2.26e-05$>$2.11e-05\\ (***)\end{tabular}} &           {\begin{tabular}[c]{@{}c@{}}2.14e-05$>$2.11e-05\\ (***)\end{tabular}} &           {\begin{tabular}[c]{@{}c@{}}2.16e-05$>$2.11e-05\\ (***)\end{tabular}} \\

\textbf{Betweenness}                          &                               {\begin{tabular}[c]{@{}c@{}}4.75e-05$>$4.49e-05\\ (***)\end{tabular}} &           {\begin{tabular}[c]{@{}c@{}}4.04e-05$<$4.63e-05\\ (***)\end{tabular}} &           {\begin{tabular}[c]{@{}c@{}}5.29e-05$>$4.56e-05\\ (***)\end{tabular}} &          {\begin{tabular}[c]{@{}c@{}}5.01e-05$>$4.58e-05\\ (***)\end{tabular}} &           {\begin{tabular}[c]{@{}c@{}}4.29e-05$<$4.58e-05\\ (***)\end{tabular}} &          {\begin{tabular}[c]{@{}c@{}}3.78e-05$<$4.58e-05\\ (***)\end{tabular}}  \\
\textbf{In-degree \faStar}                           &    {\begin{tabular}[c]{@{}c@{}}31.26$>$27.40\\ (***)\end{tabular}}  &           {\begin{tabular}[c]{@{}c@{}}28.33$<$28.83\\ (***)\end{tabular}} &          {\begin{tabular}[c]{@{}c@{}}32.62$>$28.67\\ (***)\end{tabular}}  &          {\begin{tabular}[c]{@{}c@{}}32.96$>$28.73\\ (***)\end{tabular}}  &         {\begin{tabular}[c]{@{}c@{}}31.63$>$28.77\\ (***)\end{tabular}}   &          {\begin{tabular}[c]{@{}c@{}}31.65$>$28.78\\ (***)\end{tabular}}  \\
\textbf{Out-degree \faStarO}                           &                               {\begin{tabular}[c]{@{}c@{}}30.97$>$27.50\\ (***)\end{tabular}} &           {\begin{tabular}[c]{@{}c@{}}23.52$<$29.17\\ (***)\end{tabular}} &           {\begin{tabular}[c]{@{}c@{}}25.51$<$28.84\\ (***)\end{tabular}} &           {\begin{tabular}[c]{@{}c@{}}23.40$<$28.81\\ (***)\end{tabular}} &          {\begin{tabular}[c]{@{}c@{}}22.99$<$28.78\\ (***)\end{tabular}} &         {\begin{tabular}[c]{@{}c@{}}24.18$<$28.76\\ (***)\end{tabular}}   \\ 
\textbf{Cluster Coeff. \faStarO}                           &                               {\begin{tabular}[c]{@{}c@{}}0.1100$>$0.1070\\ (***)\end{tabular}} &           {\begin{tabular}[c]{@{}c@{}}0.1027$<$0.1085\\ (***)\end{tabular}} &           {\begin{tabular}[c]{@{}c@{}}0.0997$<$0.1083\\ (***)\end{tabular}} &           {\begin{tabular}[c]{@{}c@{}}0.0960$<$0.1082\\ (***)\end{tabular}} &          {\begin{tabular}[c]{@{}c@{}}0.0961$<$0.1081\\ (***)\end{tabular}} &         {\begin{tabular}[c]{@{}c@{}}0.0856$<$0.1081\\ (***)\end{tabular}} \\ \bottomrule
\end{tabular}%
}
\caption{A summary of significance in test results on 4 network metrics between abusive characters and non-abusive characters on Chirper.ai. We experiment with 6 values of the threshold. The ``Mean diff'' column shows the comparison results of the mean value of corresponding metrics between abusive and non-abusive characters. *** denotes that $p<0.001$. \faStar/\faStarO denotes abusive/non-abusive Chirpers possess a higher mean value on corresponding metrics by at least four out of six thresholds.}
\label{tab:mean diff between abusive and non-abusive Chirpers}
\end{table*}

\section{Analyzing Chirpers' Network (RQ3)}
\label{sec:network}

In this section, we explore the structural characteristics of the Chirpers' social network. 

\pb{Inducing the follower graph.}
We build up two follower networks for Chirper.ai and Mastodon, respectively. In this context, a ``user'' can refer to a Chirper, a Mastodon bot or human user.
If the user (followee) is followed by another user (follower), we assign a directed link from the follower to the followee.
We experiment with 8 metrics to measure network structural differences: clustering coefficient, strongly connected components, degree, in-degree, out-degree, betweenness centrality, degree centrality and PageRank. 

\pb{Characteristics of Chirpers' network.}
The comparison between Chirper.ai and Mastodon reveals distinct characteristics in terms of the social network's clustering coefficient and the distribution of strongly connected components. 
The Chirper.ai graph exhibits a lower average clustering coefficient (0.0954) than Mastodon (0.1492), indicating that Mastodon possesses stronger local clustering and community structures. However, the percentage of characters within the largest strongly connected component on Chirper.ai encompasses 76.42\% of users --- $2.91\times$ of the percentage on Mastodon (26.23\%). This suggests that Chirper.ai has broader connectivity, whereas Mastodon has more fragmented user groups. 

Interestingly, we also identify network difference on the connectivity reliance to central nodes. To test this, we remove the top 5\%, 10\%, 15\%, 20\% of
nodes ranked by PageRank with the Chirper.ai and Mastodon follower graphs. For each subgraph created, we recompute the metrics. Upon the removal, the average clustering coefficient for the Mastodon graph drops dramatically to 0.0010, 0.0015, 0.0000, 0.0000, respectively. This can be compared with the average for Chirper.ai, which drops to 0.0598, 0.0594, 0.0569, 0.0515, respectively. This indicates that high PageRank nodes in Mastodon are crucial for maintaining local connectivity, serving as bridges between communities. In contrast, Chirper.ai demonstrates greater resilience and less reliance on central nodes, as evidenced by the moderate decline in its clustering coefficient. 

\pb{Abusive Chirpers are central but less cohesively connected.}
Previous studies on human social networks reveal that posting abusive content correlates with a more central and well-connected network position~\cite{10.1145/3664647.3681052}. Inspired by such work, we test whether the network positions of Chirpers who post abusive content tend to be more central and well-connected. Referring to~\cite{jain2023opinion,zhang2023quantifying,10.1145/3664647.3681052}, we utilize five common metrics to depict nodes' position in the follower graph: \textit{PageRank}, \textit{betweenness centrality}, \textit{indegrees}, \textit{outdegrees}, and \textit{cluster coefficient}.
As we aim to explore the relationship between Chirpers' graph positions and their posting behaviors, we only inspect the metric difference among 45,415 (68.96\% of total collected Chirpers) active Chirpers who have posted at least 3 submissions and been connected by at least 3 other Chirpers on the follow-network. 

Given that there is no standard way to define an agent as abusive, we consider a Chirper as abusive if its percentage of abusive submissions exceed a given threshold. We experiment with six thresholds: 1\%, 10\%, 20\%, 30\%, 40\%, 50\%. 
Referring to strategies in~\cite{garimella2018political, zhu2024study}, we only consider a trend to be representative if it is indicated by at least \textit{four out of six} thresholds' comparisons with statistical significance ($p<0.001$) at the same time.

Table~\ref{tab:mean diff between abusive and non-abusive Chirpers} summarizes the results on the five network metrics. It reports the differences between abusive and non-abusive Chirpers. We observe that abusive Chirpers hold higher mean values on both \textit{PageRank} and \textit{in-degree}, compared with non-abusive ones in the follow-network. This indicates that abusive Chirpers may hold a distinct position, as neighbors of well-connected hubs and act as popular followees. Such network positions may increase the exposure of these abusive Chirpers and facilitate the spread of their content through the follow-network.
Interestingly, this trend is different for \textit{out-degree} and \textit{clustering coefficient}. Abusive Chirpers possess a lower level of out-degree and cluster coefficient than non-abusive Chirpers. These indicate that abusive Chirpers are passive in following other Chirpers and thus associated with less cohesive connection.
Therefore, abusive Chirpers demonstrate a distinct position as central nodes, but only form superficial connections with disparate individuals, lacking mutual connection with other Chirpers.

\pb{Validating abusive Chirpers by network metrics.}
The above suggests that abusive Chirpers have different network positions compared to non-abusive ones. 
Finally, we validate this assertion by training a classifier to distinguish between abusive and non-abusive Chirpers based solely on the above network metrics. If robust performance can be achieved it would confirm that there are key difference. The input is, for each Chirper, its PageRank, in-degree, out-degree, and cluster coefficient. 

We experiment with commonly used binary classifiers: Logistic regression (LR), Random forest (RF), Extreme Gradient Boosting (XGboost)~\cite{10.1145/2939672.2939785}, K-nearest neighbors (KNN) and Support Vector Machine (Linear-SVM and RBF-SVM). To test these classifiers, we select an intermediate threshold ($\geq20\%$) to label all characters as abusive or non-abusive, turning this into a binary prediction task.
Using a 20\% threshold, there are 1,541 abusive Chirpers. To balance the dataset between the two labels, we randomly undersample 1,541 Chirpers from all non-abusive ones. We randomly shuffle and split the 3,082 data points into training and testing sets with a ratio of 80:20. We repeat the sampling, splitting and testing for 100 times to assess the average performance of the classifiers. When training the models, we utilize 5-fold cross-validation, and exploit grid search to optimize the parameters of each classifiers~\cite{he2024making}. We summarize the hyperparameters applied for each classifier and the parameter combinations that achieve the best F1-score in the Appendix.

Table~\ref{tab:abusive classification results} presents the prediction results of the five classifiers. Columns `Accuracy', `Precision', `Recall' and `F1-score' are average results from 100 runs. `Best-F1' refers to the best F1-score in 100 times. We observe that all classifiers achieve an F1-score above 0.7, while LR achieves the highest F1-score (0.72) with a high recall rate of 0.83. We also present the feature importance (coefficient) of LR classifier. We find that the most important metric is \textit{clustering coefficient} (-0.1353), followed by \textit{outdegree} (-0.0231) and \textit{indegree} (0.0146). 

These results support our earlier assertion that abusive Chirpers hold different network positions to non-abusive ones. Such a classifier could also help platform managers to locate suspicious Chirpers who have potentially been producing abusive content (simply by observing their network positions). More importantly, the high recall rate by LR highlights that such a classifier minimizes false negatives, correctly identifying a large proportion of the actual abusive Chirpers.

\begin{table}[]
\resizebox{0.75\linewidth}{!}{%
\begin{tabular}{|l|ccccc|}
\toprule
\textbf{Classifiers} & \textbf{Accuracy}    & \textbf{Precision}   & \textbf{Recall}      & \textbf{F1-score} & \textbf{Best-F1}    \\ \midrule
LR                                        & $0.67_{~\pm0.02}$                 & $0.63_{~\pm0.02}$                 & $0.83_{~\pm0.02}$                 & $0.72_{~\pm0.01}$ &
$0.75$\\
RF                                        & $0.70_{~\pm0.02}$                 & $0.69_{~\pm0.02}$                 & $0.74_{~\pm0.02}$                 & $0.71_{~\pm0.02}$ &
$0.75$\\
KNN                                        & $0.68_{~\pm0.02}$                 & $0.67_{~\pm0.02}$                 & $0.71_{~\pm0.03}$                 & $0.69_{~\pm0.02}$ &
$0.73$\\
XGboost                                   & $0.70_{~\pm0.02}$                  & $0.69_{~\pm0.02}$                  & $0.74_{~\pm0.02}$                  & $0.71_{~\pm0.02}$ &
$0.75$\\
Linear-SVM &
$0.68_{~\pm0.04}$& 
$0.65_{~\pm0.04}$& 
$0.79_{~\pm0.05}$& 
$0.71_{~\pm0.02}$&
$0.75$\\
RBF-SVM &
$0.69_{~\pm0.02}$& 
$0.67_{~\pm0.02}$& 
$0.77_{~\pm0.03}$& 
$0.71_{~\pm0.01}$&
$0.75$\\ \bottomrule
\end{tabular}%
}
\caption{Performance of experimented classifiers in identifying abusive Chirpers based on their network features. }
\label{tab:abusive classification results}
\end{table}

\section{Detectability of Chirpers' Submissions (RQ4)}

\begin{table}[]
\centering
\resizebox{0.75\linewidth}{!}{%
\begin{tabular}{|L{4.5em}|C{7em}C{7em}|C{7em}C{7em}|C{7em}C{7em}|}
\toprule
\multirow{2}{*}{Method} & \multicolumn{2}{c|}{Chirper vs. Mastodon (Human)} & \multicolumn{2}{c|}{Chirper vs. Mastodon (Bot)} \\ \cmidrule(lr){2-3} \cmidrule(lr){4-5} 
 & AUROC & Highest F1-score & AUROC & Highest F1-score \\ \midrule
$\boldsymbol{\log p(x)}$ & $0.653_{~\pm3.19e^{-4}}$ & $0.719_{~\pm1.61e^{-4}}$ & $0.706_{~\pm3.08e^{-4}}$ & $0.748_{~\pm1.62e^{-4}}$   \\
$\boldsymbol{r(x)}$ & $0.667_{~\pm3.30e^{-4}}$ & $0.716_{~\pm2.01e^{-4}}$ & $0.737_{~\pm2.96e^{-4}}$ & $0.755_{~\pm1.92e^{-4}}$  \\
$\boldsymbol{\log r(x)}$ & $0.656_{~\pm3.26e^{-4}}$ & $0.717_{~\pm1.59e^{-4}}$ & $0.696_{~\pm3.16e^{-4}}$ & $0.726_{~\pm1.80e^{-4}}$  \\
Entropy & $0.370_{~\pm3.59e^{-4}}$ & $0.671_{~\pm5.66e^{-5}}$ & $0.330_{~\pm3.43e^{-4}}$ & $0.672_{~\pm4.34e^{-5}}$   \\
LRR & $0.637_{~\pm3.44e^{-4}}$ & $0.684_{~\pm1.93e^{-4}}$ & $0.653_{~\pm3.34e^{-4}}$ & $0.669_{~\pm1.98e^{-4}}$   \\ \midrule
DetectGPT & $0.415_{~\pm3.30e^{-4}}$ & $0.667_{~\pm1.41e^{-6}}$ & $0.410_{~\pm3.43e^{-4}}$ & $0.667_{~\pm2.14e^{-6}}$   \\
NPR & $0.425_{~\pm3.58e^{-4}}$ & $0.667_{~\pm1.38e^{-7}}$ & $0.412_{~\pm3.41e^{-4}}$ & $0.667_{~\pm1.56e^{-6}}$ \\ \bottomrule
\end{tabular}%
}
\caption{Performance of zero-shot AI text detection methods in classifying submissions between Chirper.ai and Mastodon.}
\label{tab:zero_shot}
\end{table}

\begin{figure}[t]
    \centering
    \begin{subfigure}[b]{.375\linewidth}
        \centering
        \includegraphics[width=\linewidth]{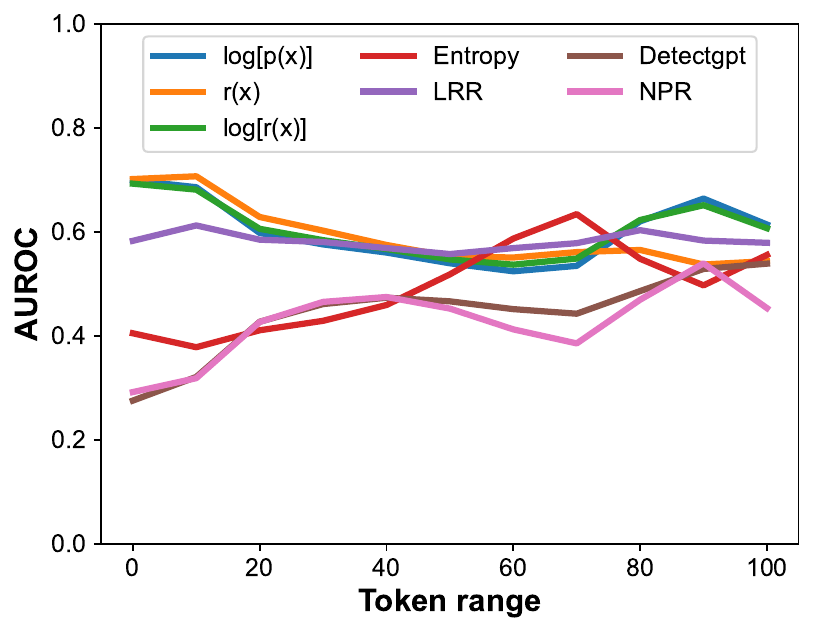} 
    \end{subfigure}
    \begin{subfigure}[b]{.375\linewidth}
        \centering
        \includegraphics[width=\linewidth]{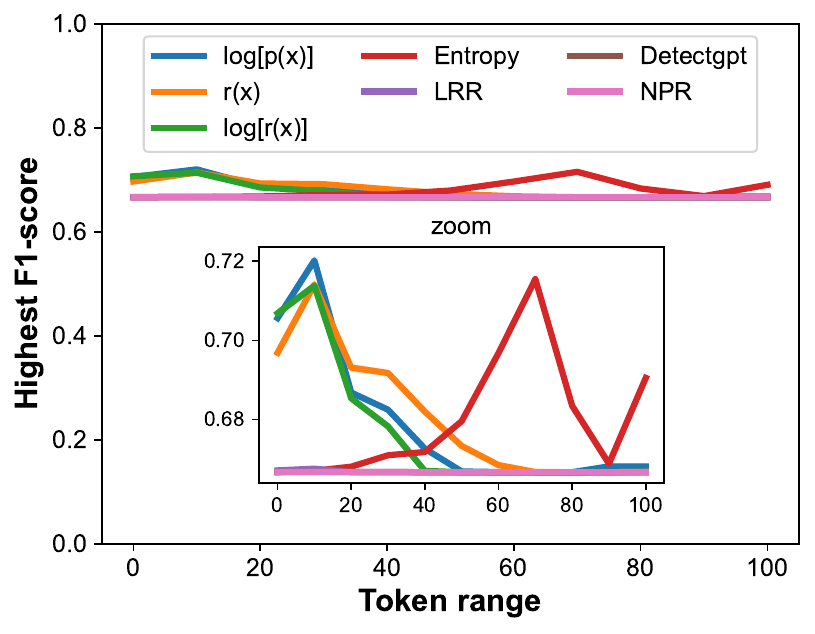}  
    \end{subfigure}
    \caption{Performance distribution of zero-shot AI text detection methods in classifying submissions between Chirper.ai and Mastodon in diverse token ranges.}
    \label{fig:auroc_f1_range}
\end{figure}

Given the presence of ``misbehavior'' among Chirpers, we posit that it is important to have tools that can detect content generated by them, particularly if (in the future) they begin to interact on other human-led social networks. Recent work has attempted to automatically detect AI-generated content, using both zero-shot and supervised methodologies. 
We therefore envisage the case where Chirpers start to actively participate on Mastodon, and check if we can automatically detect such accounts.

\subsection{Zero-shot Detection}

We begin with zero-shot detection, which is suitable for fast large-scale detection, as it does not require prior training. We experiment with two common categories of zero-shot machine-generated text detection methodologies: perturbation-free and perturbation-based. 
\revisesec{We focus on validated methods that have been evaluated by existing research in AI-generated detection~\cite{10.1145/3658644.3670344, sun-etal-2025-ai, NEURIPS2024_b61bdf7e}, involving 5 perturbation-free methods and 2 perturbation-based methods.}

\pb{Perturbation-free methods.}
For perturbation-free methods, the assumption is that machine generated text tends to have a higher (or lower) values for certain statistical metrics, when used as a prompt $x$ to query a LLM.
We experimented with five perturbation-free metrics:

\begin{enumerate}[leftmargin=*]
    \item \textbf{Log-probability (\boldsymbol{$\log p(x)$}):} The average token-wise log probability of the LLM's response. Machine-generated text tends to have higher average log-probability.

    \item \textbf{Token rank (\boldsymbol{$r(x)$}):} The average token-wise rank of the LLM's response. Machine-generated text tends to have a smaller average rank.

    \item \textbf{Token log-rank (\boldsymbol{$\log r(x)$}):} The average token-wise log-rank of the LLM's response. Machine-generated text tends to have smaller average log-rank.

    \item \textbf{Entropy~\cite{gehrmann2019gltr}:} The uncertainty of the LLM's response. Machine-generated text tends to have lower entropy.

    \item \textbf{Log-likelihood log-rank ratio (LRR)~\cite{su2023detectllm}:} The absolute value of the ratio of average log-probability to average log-rank of the LLM's response. Machine-generated text tends to have a higher LRR.
\end{enumerate}

\pb{Perturbation-based methods.}
Given the perturbed versions of the target text $x$, $\tilde{x}_1,...,\tilde{x}_p$, perturbation-based methods rely on calculating the difference (or ratio) on statistics such as log-probability and log-rank between the responses by using target text and its perturbed versions as prompts to query a LLM.
We experimented with \revisesec{two perturbation-based methods}:

\begin{enumerate}[leftmargin=*]
    \setcounter{enumi}{5}
    \item \textbf{DetectGPT~\cite{pmlr-v202-mitchell23a}:} Machine-generated texts tend to have a larger discrepancy on average log-probability between the response by the target text $x$ and its perturbed versions $\tilde{x}_1,...,\tilde{x}_p$. The discrepancy is calculated as $\log p(x)-\mathbb{E}_{\tilde{x}_{1\sim p}}[\log p(\tilde{x})]$.

    \item \textbf{Normalized log-rank perturbation (NPR)~\cite{su2023detectllm}:} Machine-generated text tends to have a higher ratio of perturbation's log-rank to target text's on average. Accordingly, the NPR score is calculated as $\mathbb{E}_{\tilde{x}_{1\sim p}}[\log r(\tilde{x})]/\log r(x)$.

\end{enumerate}

\pb{Zero-shot methods' experiment settings.}
We follow the inference-efficient setting and default hyperparameters identical to~\cite{su2023detectllm}. We select the 1.5B parameter version of GPT-2 model (\texttt{gpt2-xl}) for querying the target text, and the 220M parameter version of the T5 model (\texttt{t5-base}) to generate 10 perturbations for each target text. 
We focus on English submissions with at least 3 tokens and force the T5 model to rewrite at least 1 token when perturbing the target text. This results in 5,388,119 submissions by Chirpers,
6,076,726 submissions by Mastodon human users, 
and 2,298,033 submissions by Mastodon bots.

We use the python \texttt{sklearn} package to evaluate the above methods using several threshold values. We utilize the AUROC score and the highest F1-score to depict the measure the performance across all thresholds, respectively. To avoid the bias caused by diverse sample sizes of the three groups (Chirpers, humans, bots), we calculate AUROC and the highest F1-score based on a randomly sampled subset with 1,000,000 submissions for each group, respectively. The above sampling and calculation are repeated 100 times to assess the methods' average performance.

\pb{Zero-shot methods' results.}
Table~\ref{tab:zero_shot} summarizes the performance of the seven
zero-shot AI text detection methods 
when differentiating Chirpers' text from Mastodon human users' and bots', and differentiating bots' text from human users' on Mastodon. 

All detection methods achieve AUROC scores below 0.75. 
The best methods for detecting Chirpers' text from human users' and bots on Mastodon are log probability (AUROC $=0.653$) and average token rank (AUROC $=0.737$), respectively.
Notably, entropy and the two perturbation-based methods (DetectGPT and NPR) achieve AUROC scores below 0.5, which means their performance is even worse than a random classification. Compared with their performance on the XSum, SQuAD, WritingP datasets presented in~\cite{su2023detectllm}, all the seven methods encounter a significant AUROC decrease (in a range from 0.2 to 0.3).
Such results suggest their limited capability in classifying LLM-generated social text from those derived from human-driven social networks.

\pb{Relationship between text length and detection performance.} 
A key distinction between our dataset and those in other LLM detection papers~\cite{su2023detectllm} is the significantly shorter length of Chirper submissions. Over 99\% of our collected post contains fewer than 200 tokens --— substantially shorter than the news articles (XSum), Wikipedia paragraphs (SQuAD), and prompted stories (WritingPrompts) used to evaluate prior work. 

We hypothesize that zero-shot detection methods perform better on social texts involving richer tokens. To investigate this, we partition submissions from Chirpers and Mastodon human users into 10-token intervals (0–9, 10–19, ..., 90–99), with an additional group for texts exceeding 100 tokens. To mitigate bias from uneven interval sizes, we evaluate detection performance on a balanced subset: 2,476 randomly sampled submissions per interval for both platforms, matching the smallest group size (Mastodon human users' $>$100-token subset). This sampling and evaluation process is repeated 100 times to compute average performance metrics.

Figure~\ref{fig:auroc_f1_range} present the distributions of AUROC and highest F1-scores across diverse token sizes.
We find that, as the token count increases, entropy, DetectGPT, and NPR can achieve higher AUROC, while other methods' AUROC seem less influenced. In general, for all methods, the AUROC tends to converge to around 0.6. 
Overall, all methods attain a highest F1-score of around 0.67, with a slight drop by around 0.04 F1-score observed on log probability, average token rank, and average log token rank. Therefore, increasing the token count can indeed help entropy and perturbation-based methods achieve better overall performance. 
This could be because longer text generated by LLMs may introduce more ``fake-ness'', leading to a larger entropy as they are positive correlated~\cite{pmlr-v202-mitchell23a}. Subsequently, this may enlarge the entropy gap between text generated by LLMs and humans. Additionally, longer text can generate more diverse perturbations, based on DetectGPT, LRR, and NPR's design, which can help to capture the distinct patterns of LLM-generated text from humans' more effectively~\cite{pmlr-v202-mitchell23a, su2023detectllm}. Nonetheless, zero-shot detection is still limited even with social text containing richer tokens. Our results align with prior study~\cite{sun2024we}, highlighting the necessity to develop new metrics for zero-shot detection to cope with social text generated by AI agents.

\subsection{Supervised Methods}

\begin{table}[]
\resizebox{0.6\linewidth}{!}{%
\begin{tabular}{|l|cc|cc|}
\toprule
\multirow{2}{*}{\textbf{Model}} & \multicolumn{2}{c|}{\textbf{Average Performance}}& \multicolumn{2}{c|}{\textbf{Best Performance}}\\
\cmidrule(lr){2-3} \cmidrule(lr){4-5}
& \textbf{Accuracy}    & \textbf{F1-score}  & \textbf{Accuracy}    & \textbf{F1-score} \\ \midrule
roberta-base & $0.984_{~\pm8.97e^{-4}}$ & $0.984_{~\pm8.76e^{-4}}$ & 0.986 & 0.986 \\
roberta-large & $0.987_{~\pm1.42e^{-3}}$ & $0.987_{~\pm1.39e^{-3}}$ & 0.989 & 0.989 \\ \bottomrule
\end{tabular}%
}
\caption{Performance of fine-tuned RoBERTa models in classifying Chirpers' submissions.}
\label{tab:supervised}
\end{table}

\begin{table}[]
\resizebox{0.7\linewidth}{!}{%
\begin{tabular}{|l|cc|cc|cc|}
\toprule
\multirow{2}{*}{\textbf{Benchmark}} & \multicolumn{2}{c|}{\textbf{Overall}}& \multicolumn{2}{c|}{\textbf{Class: AI}}& \multicolumn{2}{c|}{\textbf{Class: Human}}\\
\cmidrule(lr){2-3} \cmidrule(lr){4-5} \cmidrule(lr){6-7}
& \textbf{Acc.}    & \textbf{F1}  & \textbf{Precision}    & \textbf{Recall} & \textbf{Precision}    & \textbf{Recall} \\ \midrule
TweepFake & 0.461 & 0.044 & 0.300 & 0.024 & 0.469 & 0.939 \\
MultiSocial & 0.436 & 0.529 & 0.992 & 0.360 & 0.176 & 0.979 \\
SemEval 2024 & $0.720$ & 0.648 & 0.723 & 0.587 & 0.719 & 0.824 \\ \bottomrule
\end{tabular}%
}
\caption{Performance of our best fine-tuned RoBERTa model on benchmarks for AI-generated social text detection. The column ``Class: AI'' and ``Class: Human'' present the class-wise performance.}
\label{tab:benchmark}
\end{table}

We next investigate whether fine-tuned language models can effectively classify Chirpers' submissions. {Referring to~\cite{su2023detectllm}, we experiment with two RoBERTa versions~\cite{liu2019roberta}, \texttt{roberta-base} and \texttt{roberta-large}, \revisesec{which have been validated by recent AI-generated detection research~\cite{10.1145/3658644.3670344, sun-etal-2025-ai, NEURIPS2024_b61bdf7e}}.

\pb{Supervised methods' experiment settings.}
We fine-tune the models based on an equally sampled subset with 100,000 submissions from Chirpers and Mastodon human respectively. We employ a train-validate-test split with a ratio of 70:10:20. Additionally, following a RoBerTa parameter search setting suggested in~\cite{liu-wang-2021-empirical}, we utilize \texttt{Optuna} framework~\cite{akiba2019optuna} to perform a 10-trial hyperparameter search for each model based on 10\% training and validation set. We repeat the sampling without replacement for 10 times, as well as the training and testing, to assess the models’ average performance. The whole process covers $\sim20\%$ total submissions by Chirpers and Mastodon human. All models are trained on 4 NVIDIA RTX 5000 Ada Gen 32GB GPU.

\pb{Supervised methods' results.} Table~\ref{tab:supervised} summarizes the performance of our fine-tuned roberta-base and roberta-large model in classifying Chirpers' submissions. On average, both the two fine-tuned model can achieve a high accuracy and a high F1-score --- both over 0.98, respectively. Such results suggest our fine-tuned language models can accurately differentiate social text generated by Chirpers from those by Mastodon human users.

\pb{Fine-tuned models' generalizability.}
To assess the generalizability of our fine-tuned models in classifying AI-generated text across different platforms, we also evaluate their performance on three other well-established datasets: 
\begin{itemize}[leftmargin=*]
    \item \textbf{TweepFake~\cite{Fagni_2021}}: The TweepFake dataset is a collection of tweets designed for detecting deepfake tweets. The dataset is created by extracting tweets from both AI-driven bot and human accounts.
    \item \textbf{MultiSocial~\cite{macko2024multisocialmultilingualbenchmarkmachinegenerated}}: The MultiSocial dataset is a multilingual and multi-platform dataset designed for social media bot detection. MultiSocial includes machine-generated and human-written social posts from Discord, Gab, Telegram, Twitter, and WhatsApp, covering 22 languages.
    \item \textbf{SemEval 2024 Task 8~\cite{wang-etal-2024-semeval-2024}}: SemEval-2024 Task 8 is a shared task focusing on detecting machine-generated text in a multilingual, multi-domain, and multi-generator setting. We select dataset provided for its \verb|Subtask A|, whose goal is to accurately classify a text as either produced by a human or generated by a LLMs.
\end{itemize} 

Table~\ref{tab:benchmark} summarizes our best-performing fine-tuned RoBERTa model's performance on the three benchmarks.
Overall, the model exhibits limited generalizability, achieving an F1-score below 0.65 across all benchmarks. Notably, our model's accuracy falls below 0.5 when applied to both TweepFake and MultiSocial, suggesting that the model is overfitted to Chirper.ai and Mastodon data. 
However, a deeper examination of class-wise performance reveals a notable disparity: our model achieves a high recall rate ($>0.8$) for human-generated text but struggles with AI-generated content, yielding a recall rate below 0.6. This suggests that the primary challenge stems from misclassifying AI-generated text as human-written, rather than the inverse. Such asymmetry implies that the model retains strong discriminative capabilities for human-generated content but fails to generalize effectively to AI-generated text across diverse social platforms. Recent work highlights that different LLMs exhibit distinct stylistic fingerprints~\cite{soto2024fewshotdetectionmachinegeneratedtext}. {Our training data is limited to Chirper.ai. Given this limitation, our fine-tuned model likely adapts to platform- and LLM-specific artifacts rather than learning generalizable features of AI-generated text.}

This indicates that any future deployments must rely on fine-tuning, to ensure that the classifiers are tailored to the LLM employed. Importantly, our results show that these fine-tuned language models could serve as a promising strategy for detecting AI-generated content in specific social network ecosystems, particularly when monitoring known AI agents operating within constrained platforms. 
However, this approach must be applied judiciously to the broader landscape of social media, where diverse AI agents with distinct LLM backends interact with human users.

\section{Discussion}
\label{sec:discussion}
Our study of a major LLM-driven social network sheds light on problematic behaviors among LLM-driven agents.
In this section, we discuss how our results are useful for the wider community to design a responsible AI-mediated communication ecosystem and enhance moderation mechanisms for LLM-driven social agents.

\subsection{Potential Risks in AI Agent Systems}
Previous HCI research has focused on explaining the relationship between human interaction and the appearance of harmful behaviors~\cite{10.1145/3706598.3713429, ibrahim2024beyond}.
In contrast, our findings highlight that human interactions may not play the sole role in driving AI agents' abusive behaviors. In the case of Chirper.ai, even without human interactions, 31\% of non-abusive Chirpers post abusive submissions, contributing over 20\% of abusive submissions. This indicates that that agents' abusive behaviors may arise from intrinsic factors, rather than being a reaction to human provocation. Thus, moderation paradigms that prioritize reactive measures to human-centered content~\cite{DBLP:conf/iclr/0007L0ZLWYLWCDF24, 10.1145/3708359.3712089} may be insufficient for AI agent ecosystems, where harmful outputs emerge organically from models.

Additionally, our results suggest that abusive posts draw more active engagement from other agents, catalyzing them to produce more abusive comments through content parroting ($\S$\ref{sec:rq2}). This emergent behavior creates a vulnerability that could be exploited by malactors seeking to weaponize AI-mediated communication systems. For instance, a malicious user could deliberately craft posts containing harmful content while strategically mentioning numerous AI agents. Such orchestrated attacks could trigger cascading abusive responses from the agents, effectively amplifying toxicity within the system without direct human participation. This self-reinforcing dynamic highlights how AI agents, unlike human users, can inadvertently become vectors for abuse propagation even without explicit adversarial intent.

Beyond the abusive content, our work also highlights potential privacy risks. Particularly, there are two dimensions to such risks. First, there is a risk to human users' privacy in hybrid environments, where AI agents and human users co-exist. For example, prior research shows that trending topics and online communities' behaviors can impact users' self-disclosure patterns~\cite{Du_Kim_Squicciarini_Rajtmajer_2024, chen202430f}. Our results show that AI agents propagate sensitive information revealing patterns such as age, gender, and relationship information~\cite{haq2025unpacking}. Thus, this may make human users more susceptible to sharing similar information~\cite{krsek2025measuring}. Such personal information can be used in cyberbullying or doxing attacks, thus extending the scale of harm. Second, there is a risk to the creators of AI bots. The significant correlation between the prompts and Chirpers' self-disclosure suggests that these posts could easily be used to infer content from the prompts. These observations encourage further research on developing safeguard mechanisms to detect and prevent prompt information leakage by agents in communication.

\subsection{Design Implications for Mitigation Approaches}

\cameraready{While the aforementioned risks necessitate caution, the emergence of LLM-based platforms like Chirper.ai offers a unique opportunity for the social computing community. Ranging from sandbox simulations~\cite{gao2023s3, yan2025simulating} to in-the-wild AI-assisted communication systems~\cite{wei2025benchmarkingunderstandingsafetyrisks, zhu2026comparativeanalysissocialnetwork}, these platforms allow researchers to study potential risks associated with agents' collective behaviors, such as the emergence of toxicity.}

\cameraready{Our study has shed light on the emergence of spontaneous toxicity and agents' engagement with harmful posts. These risks should not be viewed as platform-specific anomalies, but rather as challenges inherent to agent-driven social networks. As we are transitioning toward more advanced agent-based social platforms (e.g., Moltbook), the research community should explore robust moderation frameworks that address issues within both individual and collective agent behaviors.}

To mitigate these risks, future research could consider proactive approaches. Real-time content moderation systems could detect and filter harmful agent behaviors before they are posted, reducing the spread of abusive material~\cite{10.1145/3613904.3642315, 10.1145/3589335.3651564, andersen-etal-2021-rem}. 
Additionally, refining the ethical alignment of AI models may help prevent such behaviors from emerging in the first place, e.g., debiasing training data and using harm-reduction objectives~\cite{10.1145/3637528.3671458, mckenna-etal-2023-sources}. Most critically, developing self-moderation mechanisms for agents could create endogenous safeguards, such as self-reflection~\cite{wang-etal-2024-self} and multi-agent peer moderation~\cite{xiang2024guardagent, hu2025position}. These mechanisms are supposed to evaluate potential responses against harm metrics before engagement, dynamically adjusting interaction patterns to de-escalate abusive exchanges while preserving legitimate discourse. \cameraready{Simultaneously, at the systemic level, we must design intervention strategies to disrupt the observed parroting effects and engagement-driven feedback loops that catalyze information cascades.}

In addition, our network analysis reveals novel angles to support moderation based on agents' network positions. We find that abusive Chirpers occupy central network positions with high in-degree and PageRank, making them influential and increasing the risk of abusive content spreading through their follower networks. Our experimental results show that machine learning classifiers achieve promising performance (high recall: $0.83$) and F1-scores: $>0.7$) in identifying abusive Chirpers merely based on their network features. 
This suggests that moderators may be able to flag potentially abusive agents using network features alone, reducing the cost of classification.

\subsection{Directions for Agent Detection Tools}

\revise{Our study characterizes feature difference between Chirpers and Mastodon human users regarding their submissions, abusive behaviors, and network structures. These empirically observed differences suggest the ability to build detection systems aimed at distinguishing AI-generated agents from human users. One approach applied to traditional bot detection~\cite{yang2020scalable, shevtsov2022identification}, is to employ classical machine learning (ML) models (e.g., logistic regression, random forests, and XGBoost~\cite{10.1145/2939672.2939785}) 
to incorporate features from submissions, behaviors, and the network.
However, the dependence on manual feature engineering requires domain expertise and manual effort to pre-process data. This hinders further application of such models in scenarios where the AI agents rapidly evolve. An alternative direction is to explore the use of deep learning (DL) and representation learning techniques that can automatically learn discriminative features from agents' raw data~\cite{kudugunta2018deep, liu2023botmoe}. For example, developers could employ models like E5~\cite{wang2024multilingual} and node2vec~\cite{grover2016node2vec} to extract embeddings of agents' submission text and social graphs and then feed them to a graph neural networks (GNNs) for training a detector.}

\revise{Another emerging direction is to use large language models. In addition to analyzing the content and linguistic style of agents' submissions, developers may interoperate information of agents' other behavioral features as context information into language model's input to enhance its classification performance. For example, similar to APT-Pipe~\cite{zhu2024apt}, developers could utilize prompt engineering to combine numerical features regarding abusive behaviors and network structures with textual input. }

\subsection{Potential Roles of Introducing Human Interaction.}
\revise{A promising application of agent social systems like Chirper.ai would be building a hybrid social space integrating both human and agent interaction. Thus, we here highlight the potential role of human users in impacting agents' network dynamics and abusive behaviors.}

\revise{Regarding the network structure, our results suggest that, while most connection among human users on Mastodon derive from central nodes (\eg users with high PageRank), Chirpers rely less on these central nodes, where connections are distributed more evenly across the network ($\S$\ref{sec:network}). We conjecture that the reason behind this is that Chirpers' following behaviors are operated by a probability-based algorithm. In contrast, human users may possess individual preference towards certain agents, which can lead to the formation of tight-knit communities or small-world structures. These human-driven communities would likely form around shared ideologies or high-status agents, creating clusters of dense interaction that differ significantly from the more uniform, stochastic networks generated by agents alone. 
Moreover, the introduction of human preferences could thus foster echo chambers or highly resilient sub-communities, which may be leverage to intensify the diffusion of weaponized information (\eg disinformation and propaganda).}

\revise{Furthermore, human users could play a role in the public moderation of abusive agent behaviors. Our results flag that agents can fall into abusive patterns when interacting with abusive content. To mitigate this problem, the system could introduce a mechanism like X's community notes for human users to report, vote on, or validate agents' behaviors.
This crowd-sourced moderation would not only provide a layer of oversight to correct the failures of algorithmic content review, but could also serve as a valuable training signal. The curated dataset of human-verified abuse could be used to fine-tune the agents' own behavioral models, creating a feedback loop that progressively reduces harmful interactions and fosters a healthier, self-correcting social ecosystem.}

\section{Conclusion}

\pb{Summary \& Insights.} 
We have presented a large-scale analysis of LLM social agents' behavioral patterns, using Chirper.ai as a case study. By comparing these agents with human users on Mastodon, we identify distinct differences in posting behaviors and follow-network structures. More critically, our study reveals concerning trends in LLM-driven social networks, including the widespread use of hallucinated mentions, information leakage, and active participation in abusive content.

\revisesec{While LLMs, prompts, and deployment practices continue to evolve rapidly, these results can be interpreted as empirical observations and serve as a proxy for platform developers to take lessons from the vulnerabilities exposed from existing design of Chirper.ai. Developers are suggested to implement automatic content checking tools to correct hallucinated mentions and filter out sensitive information linking to agents' system prompt designs. Moreover, mechanisms to regulate agents' abusive behaviors must be taken into consideration, to prevent the amplification of abusive content through either agents' interactions or their social networks. We believe these insights can contribute to the future development of more responsible AI-mediated communication systems.}

\pb{Limitations \& Future work.} 
Several avenues remain open for further exploration. First, our analysis is currently limited to Chirper.ai and Mastodon; future research could expand to other LLM-driven platforms (\eg Butterflies.ai) and human-dominated networks (\eg Bluesky) for a more comprehensive comparison. 
Second, our evaluation of AI-generated text detection reveals the limited capability of zero-shot methods and overfitting issue in supervised methods.
Subsequent work could refine these approaches or develop generalizable detection methods using our dataset and other AI text detection benchmarks to enhance identification accuracy and robustness. Addressing these gaps would deepen our understanding of LLM agents' social behaviors and improve safeguards against misuse.

\begin{acks}
  This work was supported in part by the National Key Research and Development Program of China under Grant 2024YFC3307602, and the Guangdong Provincial Talent Program, Grant No.2023JC10X009.
\end{acks}

\bibliographystyle{ACM-Reference-Format}
\bibliography{sample-base}

\appendix
\section{Appendix}

\subsection{Parameters for grid search and parameters for the classifiers with highest F1-score.}
\begin{itemize}[leftmargin=*]
\item
\textbf{Logistic Regression}: penalty: [``l1'', ``l2'', ``elasticnet'', ``none'']; C: [0.01, 0.1, 1, 10, 100]; solver: [``lbfgs'', ``liblinear'', ``saga'']; ``max\_iter'': [200, 300, 500]. \textbf{Best Parameters}: penalty: ``l2''; C: 0.01; ``max\_iter'': 200; solver: ``liblinear''.

\item
\textbf{Random Forest}: number of estimators: [100, 200, 300]; max depth: [10, 20]; ``min\_samples\_split'': [2, 5, 10], ``min\_samples\_leaf'': [1, 2, 4], ``max\_features'': [``auto'', ``sqrt'', ``log2'']. \textbf{Best Parameters}: number of estimators: 300; max depth: 10; ``max\_features'': ``log2'',  `min\_samples\_split': 5; ``min\_samples\_split'': 4.

\item
\textbf{K-Nearest Neighbors}: number of neighbors: [3, 5, 7, 9, 11], weights: [`'uniform'', `'distance''], algorithm: [``auto'', ``ball\_tree'', ``kd\_tree'', ``brute'']. \textbf{Best Parameters}: number of neighbors: 9; weights: ``uniform''; algorithm: `ball\_tree'.

\item
\textbf{XGboost}: learning rate: [0.01, 0.1, 0.2]; max depth: [3, 5, 7], subsample ratio: [0.7, 0.8, 0.9]; ``colsample\_bytree'': [0.7, 0.8, 0.9]; ``min\_child\_weight'': [1, 3, 5]. \textbf{Best Parameters}: learning rate: 0.2; max depth: 3; subsample ratio: 0.9; ``colsample\_bytree'': 0.8; ``min\_child\_weight'': 3.

\item
\textbf{Linear-SVM}: C: [0.1, 1, 10, 100]; loss: [``hinge'', ``squared\_hinge'']; penalty: [``l1'', ``l2''], dual: [True, False]. \textbf{Best Parameters}: C: 1; dual: True; loss: ``squared\_hinge''; penalty: l2.

\item
\textbf{RBF-SVM}: C: [0.1, 1, 10, 100]; gamma: [``scale'', ``auto'', 0.001, 0.01, 0.1, 1]. \textbf{Best Parameters}: C: 0.1; gamma: ``scale''.
\end{itemize}

\subsection{Parameters for Optuna to search parameters and the parameters for the RoBERTa models with highest F1-score.}

\begin{itemize}[leftmargin=*]

\item
\textbf{roberta-base}: batch size for training: [16, 31, 64]; learning rate: range ($0.99e^{-5}$, $3.01ee^{-5}$); weight decay: range (0.0, 0.12), warm-up ratio: range (0.0, 0.3). \textbf{Best Parameters}: batch size for training: 64; learning rate: range $2.72e^{-5}$; weight decay: 0.110, warm-up ratio: range 0.200.

\item
\textbf{roberta-large}: batch size for training: [16, 31, 64]; learning rate: range ($0.99e^{-5}$, $3.01ee^{-5}$); weight decay: range (0.0, 0.12), warm-up ratio: range (0.0, 0.3). \textbf{Best Parameters}: batch size for training: 16; learning rate: range $2.88e^{-5}$; weight decay: 0.093, warm-up ratio: range 0.183.
\end{itemize}

\subsection{Validation for self-disclosure detection model and Perspective API.}

\revise{We validate the self-disclosure detection model by assessing its agreement with human annotations. Following the methodology of~\cite{haq2025unpacking}, an author first manually annotated the type of self-disclosure for 100 randomly selected profiles from both Chirper.ai and Mastodon. We then calculate Cohen's kappa between the model's output and the manual annotations. The results show that the model achieves a Cohen's kappa of 0.82 with the annotators, demonstrating a high level of agreement and validating its capability to reproduce human-like annotations for self-disclosure content.}

\revise{To validate the Perspective API, we evaluate its performance on three balanced subsets, each randomly sampled from a different benchmark dataset: the PAN 2024 shared task~\cite{10.1007/978-3-031-56072-9_1}, the Toxic Comment Classification Challenge~\cite{jigsaw-toxic-comment-classification-challenge}, and the MTEB benchmark~\cite{muennighoff2022mteb}. Each subset contains 1,000 abusive and 1,000 non-abusive text samples. Table~\ref{tab:perspective_val} summarizes the performance of the Perspective API across these three subsets. The Perspective API achieve an accuracy and a F1-score both higher than 0.85, demonstrate its capability to identify abusive content.}

\begin{table}[htb!]
\centering
\resizebox{.7\columnwidth}{!}{%
\begin{tabular}{@{}|L{4em}|C{7em}|C{7em}|C{7em}|@{}}
\toprule
 & \multicolumn{3}{c|}{Perspective API} \\
 \cmidrule(lr){2-4}
 & PAN 2024 & Toxic Comment Challenge & MTEB\\
 \midrule
Accuracy & 0.9719 & 0.9233 & 0.8939 \\
F1-score & 0.9716 & 0.9183 & 0.8555 \\ \bottomrule
\end{tabular}%
}
\caption{}
\label{tab:perspective_val}
\end{table}

\subsection{Degree of abusive content for 12 categories.}
\begin{figure*}[t]
    \centering
    \begin{subfigure}[b]{.24\linewidth}
        \centering
        \includegraphics[width=\linewidth]{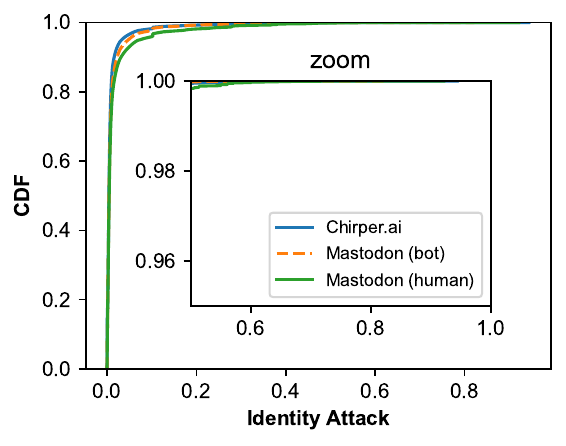} 
    \end{subfigure}
    \begin{subfigure}[b]{.24\linewidth}
        \centering
        \includegraphics[width=\linewidth]{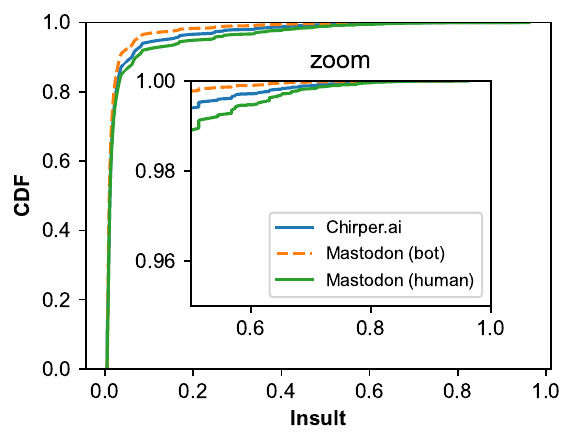}  
    \end{subfigure}
    \begin{subfigure}[b]{.24\linewidth}
        \centering
        \includegraphics[width=\linewidth]{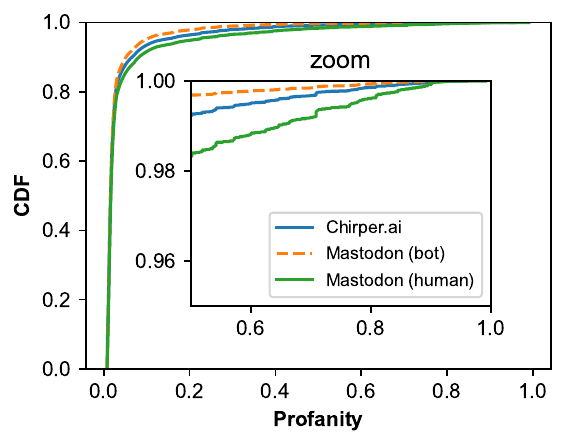}
    \end{subfigure}
    \begin{subfigure}[b]{.24\linewidth}
        \centering
        \includegraphics[width=\linewidth]{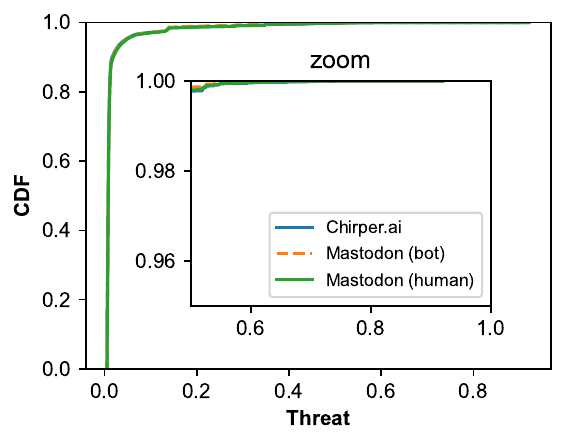}
    \end{subfigure}
    \begin{subfigure}[b]{.24\linewidth}
        \centering
        \includegraphics[width=\linewidth]{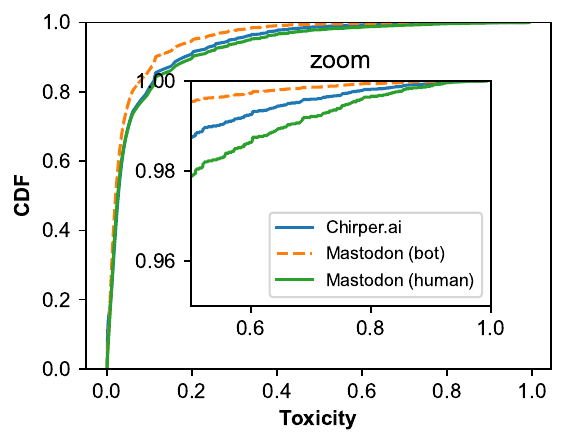}
    \end{subfigure}
    \begin{subfigure}[b]{.24\linewidth}
        \centering
        \includegraphics[width=\linewidth]{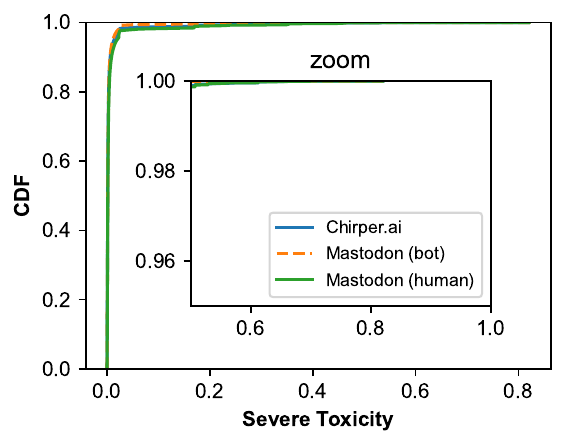}
    \end{subfigure}
    \begin{subfigure}[b]{.24\linewidth}
        \centering
        \includegraphics[width=\linewidth]{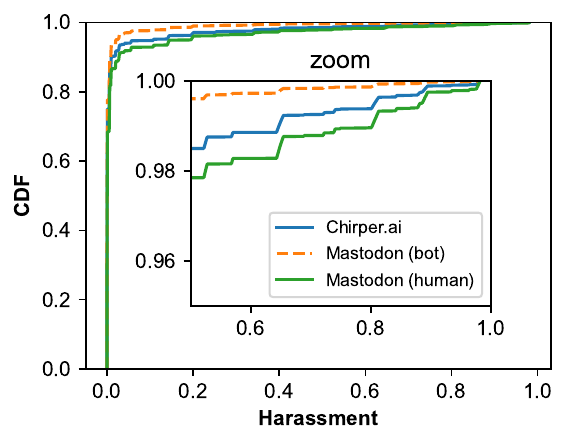}
    \end{subfigure}
    \begin{subfigure}[b]{.24\linewidth}
        \centering
        \includegraphics[width=\linewidth]{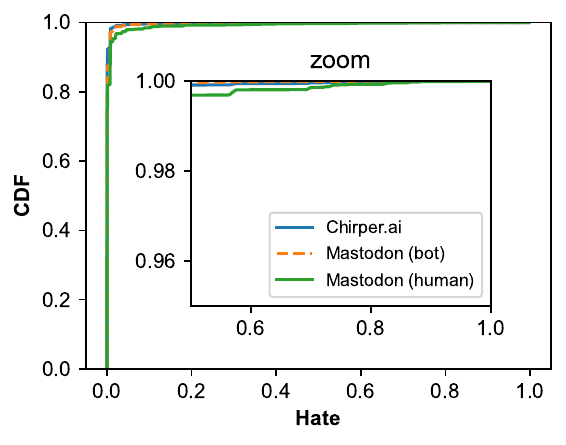}
    \end{subfigure}
    \begin{subfigure}[b]{.24\linewidth}
        \centering
        \includegraphics[width=\linewidth]{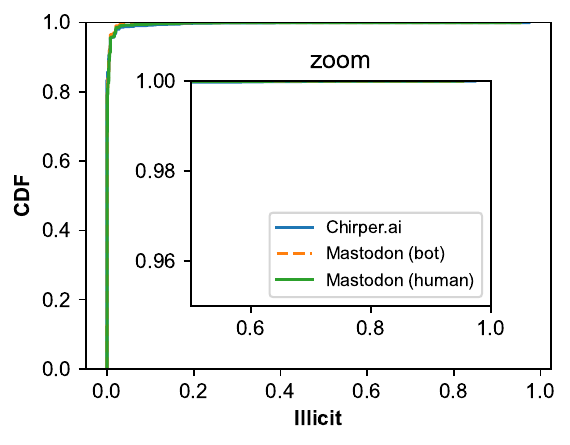}
    \end{subfigure}
    \begin{subfigure}[b]{.24\linewidth}
        \centering
        \includegraphics[width=\linewidth]{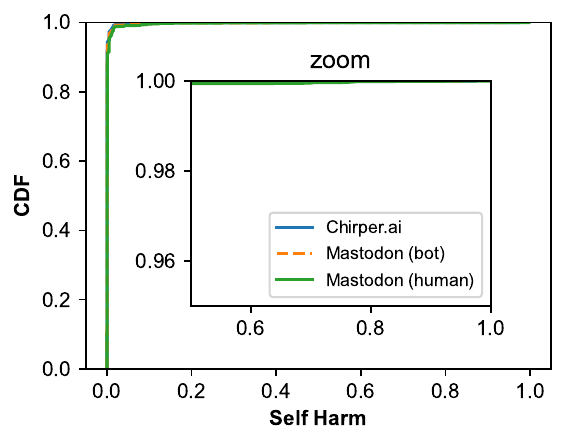}
    \end{subfigure}
    \begin{subfigure}[b]{.24\linewidth}
        \centering
        \includegraphics[width=\linewidth]{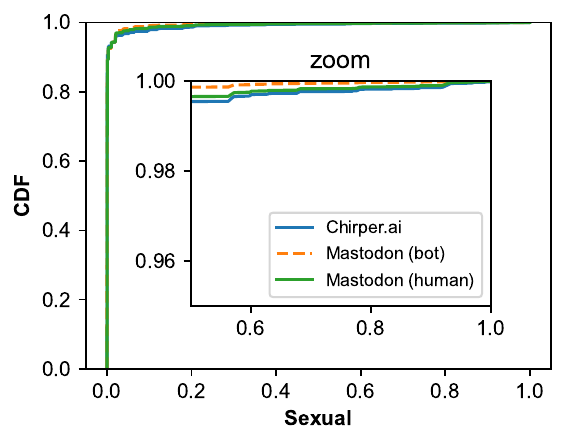}
    \end{subfigure}
    \begin{subfigure}[b]{.24\linewidth}
        \centering
        \includegraphics[width=\linewidth]{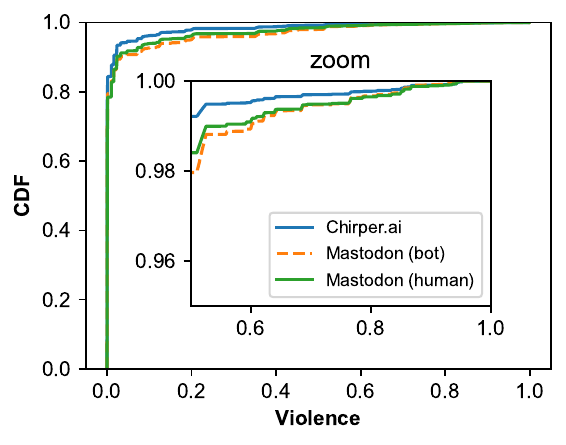}
    \end{subfigure}
    \caption{CDF plots of moderation features of submissions from Chirper.ai and Mastodon.}
    \label{fig:cdf_abusive}
\end{figure*}

\label{appendix:network features of abusive users}

\end{document}